\documentclass[english,twocolumn,prx,superscriptaddress,longbibliography]{revtex4-2}
\usepackage[T1]{fontenc}
\usepackage[utf8]{inputenc}
\usepackage{color}

\usepackage[colorlinks=true,linkcolor=blue,urlcolor=blue,citecolor=blue,anchorcolor=blue]{hyperref}
\makeatletter
\usepackage{amsmath}
\usepackage{subfigure}
\usepackage{graphicx, mathtools}
\usepackage[normalem]{ulem}

\makeatother

\usepackage{babel}
\begin{document}
\newcommand{\be}{\begin{equation}}
\newcommand{\ee}{\end{equation}}
\newcommand{\bn}{\begin{eqnarray}}
\newcommand{\en}{\end{eqnarray}}
\newcommand{\ii}{\'{\i}}
\newcommand{\ca}{\c c\~a}
\newcommand{\uc}{\uppercase}
\newcommand{\tb}{\textbf}
\newcommand{\bw}{\begin{widetext}}
\newcommand{\ew}{\end{widetext}}

\title{Anomalous Quantum Criticality at a Continuous Metal-Insulator Transition}

\author{M. S. Laad}
\email{mslaad@imsc.res.in}
\affiliation{The Institute of Mathematical Sciences, Taramani, Chennai 600113, India, and\\ 
Homi Bhabha National Institute, Anushakti Nagar, Trombay, Mumbai 400085, India}

\author{Prosenjit Haldar}
\email{prosenjit.phy@faculty.nita.ac.in}
\affiliation{%
Department of Physics, National Institute of Technology Agartala, Jirania, West Tripura 799046, India
}%

\date{\rm\today}

\begin{abstract}
   The Falicov-Kimball model (FKM) is long known to be the simplest model of correlated fermions exhibiting a novel Mott-like quantum critical point (QCP) assocaited with a {\it continuous} MIT in dimensions $D \geq 3$.  It is also known to be isomorphic to an {\it annealed} binary-alloy disorder model.
   Notwithstanding extensive numerical studies for
 the FKM, analytic insight into the microscopic processes spawning novel Mott-like quantum criticality is scarce.  Here, we develop a fully analytic theory for the
	Mott-like quantum criticality in the FKM on a hierarchical 
Cayley tree (Bethe lattice) by utilizing a single input from a 2-site cluster-dynamical mean-field theory (CDMFT).  We find that density fluctuation modes acquire anomalous dimensions, originating from infra-red power-law singular cluster self-energies.  Interestingly, we uncover, at $T=0$, that this {\it sub-diffusive} metal with 
glassy dynamics separating a weakly ergodic metal from a non-ergodic insulator shrinks to a single point, namely the Mott-like QCP, at least on the Bethe lattice.  We detail the
consequences of this anomalous quantum criticality for a range of thermal and dynamical
responses in a variety of physical systems that can be effectively modelled by the FKM.
\begin{description}
\item[Keywords]
Strongly Correlated Electronic System, Mott Insulator, Quantum criticality
\end{description}
\end{abstract}

\maketitle

%\vspace{1.0cm}
\section{Introduction}
   Correlation-driven Metal-Insulator Transitions (MITs) are one of the basic phenomena underpinning the anomalous physical responses of strongly correlated electron systems (SCES), and continue to foment novel developments in quantum matter.  Dynamical mean-field theory (DMFT) and its cluster extensions are now 
 poised to describe unconventional metallic and ordered states that frequently 
 occur near the MIT in physical systems of intense interest~\cite{Imada,Kotliar}.  These studies provide increasing insight into anomalous metallic states that emerge at the boundaries of such quantum phase transitions characterized 
 by full or partial Mott localization of carriers.  In many cases, the critical 
 end-point (CEP) of the line of first-order MITs can be driven down to low $T$, e.g, in organics.  DMFT studies show~\cite{terletska, kanoda} novel Mott quantum-critical behavior {\it above} this Mott CEP, at low $T_{CEP}$.  Interestingly, it is found that $T_{CEP}> T^{*}$, the lattice coherence scale below which a 
 correlated Landau Fermi liquid (LFL) metal obtains, so that the ``quantum critical features'' seen, {\it e.g}, in organics are associated with an anomalous metal without Landau Fermi liquid (LFL) quasiparticles (for e.g, such a metal shows a linear-in-$T$ resistivity, reminiscent of a ``strange'' metal~\cite{terletska}).

  Possibility of tuning the CEP to $T=0$ is thus of intense interest in context
  of searching for the elusive quantum-critical nature of the MIT.   
  While extensive work continues to address this issue in the context of the 
  Hubbard model~\cite{Vojta} it is long known that a disorder-driven MIT shows a genuine quantum critical {\it point} (QCP)~\cite{Ohtsuki}.  
  This class of models bear a close relation to the simplest model of correlated electrons on a $D$-dimensional lattice, the Falicov-Kimball model (FKM).  In symmetry-unbroken cases, the FKM also hosts a ``Mott-like'' QCP in high-$D$, associated with a continuous band-splitting type of metal-insulator transition, in DMFT as well as its cluster extensions~\cite{hubbard, vollhardt,vollhardt_err,krishnamurthy, freericks,haldar-laad-hassan}.  
The FKM can also be viewed as a binary-alloy {\it annealed} disorder model 
for the itinerant subsystem~\cite{Zonda_2012,Zonda_2009}.  It is also viewable as a simplified version of the one-band Hubbard model, where electrons with one (say $\downarrow$) spin component are immobile.  At least in high dimensions, and above the lattice coherence scale, $T^{*}$,
the Hubbard model can be considered as two effective, self-consistently coupled FK models
when thermal fluctuations destabilize the lattice Kondo scale~\cite{Edwards} (this is the
so-called {\it static} spin fluctuation limit, and corresponds to an {\it effective} suppression of coherent hopping of one spin species).  The second-order transition at the CEP of the line of first-order metal-insulator transition in the organics lies above $T^{*}$, so it is an interesting issue as to whether the anomalous criticality in the vicinity of $T_{CEP}$ can be related to two, self-consistently coupled FK models.  Such models have indeed been 
considered as effective low-energy models for CMR manganites~\cite{Ramakrishnan} and for the adiabatic limit of Holstein model(s) of electrons coupled to optic
 phonons~\cite{Majumdar,Ciuchi,PRB_Thoss_2019,PRB_Zonda_2020}.  While the analogy to the 
binary disorder model suggests an Anderson metal-insulator transition (MIT) 
driven by (annealed) disorder, the FKM also exhibits a {\it non-Anderson} MIT, 
characterized by onset of critical behavior in the {\it disorder averaged} density-of-states (DOS) at the MIT.  Interestingly, recent work has investigated a 
related issue~\cite{Radzihovsky,Radzihovsky1} in the context of disorder-driven MIT in 
Weyl semimetals, among others.  We have recently investigated anomalous Mott-like quantum criticality and it's consequences for various electrical and thermal 
responses in detail in this latter context in a series 
of papers in a simplest, $2$-site cluster-DMFT (a cavity approximation exact to $O(1/D)$) for the FKM~\cite{haldar-laad-hassan}.  
Notwithstanding the large amount of numerical work in this context, analytic 
progress has been slow, limiting detailed understanding 
of the ``Mott-like'' QCP as well as the microscopic processes that spawn the 
observed anomalous responses near the continuous MIT~\cite{PRB_ML1,PRR_ML2}.  Consequences of 
the critical slowing down in dynamics near the MIT, like onset of anomalous 
(sub)-diffusion, onset of glassiness and breakdown of ergodicity are issues 
that have been a focus of intense interest in recent studies of disordered, 
{\it interacting} models (mostly in $D=1$), but such effects in higher dimension $D>1$ are lacking, and call for systematic investigation.

\vspace{1.0cm}

  The FKM, defined by the Hamiltonian

\be
H = -t\sum_{<i,j>}(c_{i}^{\dag}c_{j}+h.c) + U\sum_{i}n_{ic}n_{id} -\mu\sum_{i}n_{ic}
\ee
possesses en exact local U$(1)$ gauge symmetry, since $[n_{id},H]=0$ for all
$i$.  $H$ encapsulates a curious dual aspect: while the localized 
($d$) fermions provide an annealed disorder for the conduction ($c$) fermions 
(thus, the FKM for the itinerant $c$-fermions is isomorphic to an Anderson 
disorder model, but {\it not} for the immobile $d$-electrons), a 
$d$-fermion mediated statistical interaction between $c$-fermions also emerges.  This is generally true, but most easily seen in the large-$U$ limit of the FKM, where elimination of site double-occupancy leads to a $t-J_{z}$ model, given by

\be
H_{t-J_{z}}=-t\sum_{<i,j>}(X_{i}^{\dag}X_{j}+h.c) + J_{z}\sum_{<i,j>}(S_{i}^{z}S_{j}^{z}-\frac{1}{4}n_{i}n_{j})
\ee
with $J_{z}=4t^{2}/U$, $X_{i\sigma}=(1-n_{i,d})c_{i}$ and $S_{i}^{z}=(n_{ic}-n_{id})/2, n_{i}=(n_{ic}+n_{id})$.
Clearly, both, the hopping (because $X_{i}^{\dag}X_{j}=(1-n_{id})c_{i}^{\dag}c_{j}(1-n_{dj})=P_{i,j}c_{i}^{\dag}c_{j}$, wherein the $d$-fermion 
configuration-dependent hopping is manifest) {\it and} the interaction (magnitude $J_{z}$, involving a $c-d$ interaction between neighbor sites) depend on the 
random $d$-fermion configuration.  Both are now randomly fluctuating functions 
that are slaved to the random $d$-fermion configuration.  Explicitly,  

\be
H_{t-J_{z}}=-t\sum_{<i,j>}P_{i,j}(c_{i}^{\dag}c_{j}+h.c) - (J_{z}/2)\sum_{<i,j>}(n_{ic}n_{jd}+n_{id}n_{jc})
\ee

where $P_{i,j}=(1-n_{i,d})(1-n_{j,d})$.
These features are intrinsic to the FKM at arbitrary coupling, and their 
dynamical effects are captured by an $O(1/D)$ approximation within cDMFT (but 
not within single-site DMFT).  Thus, quite generally, the FKM may 
be understood as a spinless model where non-local $c,d$-fermion interactions 
compete with a $d$-fermion configuration-dependent random hopping.  The main 
difference between an Anderson disorder model with off-diagonal hopping and Eq.(3) is that the $d$-fermions have non-trivial dynamics akin to a ``sudden, local
 quantum quench'' as viewed from the $c$-fermion perspective.

\vspace{0.5cm}

   Interestingly, {\it precisely} such models have been of great current interest in many-body localization (MBL) lore.  
 They are also central to ``quantum disentangled liquids''~\cite{MPAFisher}, wherein one has a mixture of light and heavy quantum entities: while the heavy 
particles are fully thermalized, the light particles may, under appropriate 
conditions, {\it not} reach thermalization.  This would imply invalidation of 
one of the central assumptions of statistical mechanics.  In a microcanonical 
ensemble, the very notion of temperature, $T$, arises from counting the total 
number of configurations with total energy between $E$ and $E+dE$ and 
equating it's (logarithmic) energy derivative with $\beta=1/k_{B}T$.  
''Thermal equilibrium'' is reached when the system passes through all these 
configurations over a time scale identified with the equilibration time 
($t_{eq}$).  In a strongly disordered (or glassy) system, 
this configurational mixing leading to equilibration takes place over 
 very long time scales that depend on the degree of disorder (or glassiness).  
 For times $t<t_{eq}$, the very idea of temperature does not exist, and ergodicity ``breaks down''.  When $t_{eq}=\infty$, the system can never equilibrate. 
 
   The disorder-driven metal-insulator transition is characterized by the 
   absence of diffusion and onset of non-ergodicity, as noticed very early on by Anderson himself~\cite{Anderson}.  In the more recent MBL context, it is argued that a non-ergodic {\it bad} metal~\cite{Altshuler,Huse,Huse1,Prelov} with anomalous, 
 sub-diffusive responses intervenes between an ergodic, diffusive metal and a non-ergodic insulator.  The sub-diffusive bad metal shows features reminiscent of quantum Griffiths phases, and glassy dynamics.  Given that the FKM can be viewed as a system where the ``heavy'' particles provide an {\it annealed} disorder for the 
 ``light'' electrons, and thus as a lattice model for the QDL, one may wonder about relevance of these issues in the context of the band-splitting type 
 of MIT in this case.  
 
    In this article, we investigate these issues of $(i)$ subdiffusive metallicity in the FKM and $(ii)$ possible onset of non-ergodicity, in particular, whether an intermediate non-ergodic bad metal phase obtains, or whether it shrinks to a single point, coinciding with the MIT.  While we indeed find a range 
of features that are reminiscent of MBL a low but finite $T$ in our case, we do {\it not} find a 
non-ergodic ``metal'' {\it phase} separating the diffusive metal from a non-ergodic insulator {\bf at} $T=0$.  Remarkably, however, we find breakdown of ergodicity {\it precisely} at the quantum critical point separating the diffusive metal from a non-ergodic ``Mott-like'' insulator at $T=0$: thus, the conjectured subdiffusive non-ergodic metallic phase shrinks to a single quantum critical point.  Such behavior is expected at an Anderson transition in $D>2$~\cite{Tomasi}, 
 so our result showing a similar feature at the Mott-like QCP is very unusual and of broader import.  Moreover, since a random binary alloy model is a proxy of the many-body localization (MBL) transition on the Fock space of generic interacting systems (Logan), our cavity approximation could open the route to study subdiffusive dynamics in Fock space, especially since recent studies~\cite{Luitz} indicate that the subdiffusive dynamics can be consistent with weak ergodicity.
In this paper, we consider only the {\it non Anderson} MI transition in 
the FKM, and discuss effects of Anderson localization within a two-site CDMFT in brief later (see Pages 12, 13).

\vspace{0.5cm}

   Here, using a single input from our recent 2-site CDMFT studies for the FKM 
   as a starting point, we undertake an analytic approach to unearth $(i)$ the 
   nature of quantum criticality at the non-Anderson MIT in the FK model and $(ii)$ its manifestations in a wide range of physical responses across the MIT.
  We begin with a {\it random} FKM, with a binary-alloy $d$-fermion distribution: $P(U_{i})=(1-x)\delta(U_{i}) + x\delta(U_{i}-U)$: this corresponds to ignoring
 the symmetry-broken state(s) like electron crystallization/charge density wave
that (independent of dimensionality $D$) are known to be the true ground states
 of the FKM.  In a cDMFT framework, the ($O(1/D)$) ``alloy'' correlations, 
given by $f_{ij}=\langle n_{i,d}n_{j,d}\rangle - \langle n_{i,d}\rangle\langle n_{j,d}\rangle$ with $i,j$ nearest neigbors, explicitly enter the propagators and
 the self-energies.  In CMR manganites, for e.g, the $f_{ij}$ can be identified
 with Jahn-Teller short-range order, 
the importance of which is seen in an analysis of the pair distribution function in EXAFS data~\cite{Attwood}.  Such situations are widespread in doped 
transition-metal oxides and disordered alloys (the latter have 
been extensively studied, mostly using single-site coherent potential approximation (CPA)~\cite{CPA}).  Specifically, we use the CDMFT self-energy, 
$\Sigma({\bf K},\omega)$ (here, ${\bf K}$ are the intra-cluster momenta, equal 
to $(0,0),(\pi,\pi)$ in our two-site cluster) as the sole input from numerics 
to construct a self-consistent theory for the continuous MIT in the 
FKM.  CPA has been used earlier~\cite{kroha} to study Anderson localization.   
Our work is an extension that exactly captures dynamical correlation effects 
to order $1/D$, and though the MIT  is continuous, as in DMFT, the self
energy has a very different behavior near the MIT~\cite{haldar-laad-hassan}.
  Here, we detail specific manifestations of strong non-local dynamical 
correlations {\it beyond} DMFT on the novel Mott-like QCP associated with the 
complex interplay between strong {\it annealed} disorder and carrier itinerance.
 
It is remarkable that the model is amenable to an (almost) analytic solution, 
both in DMFT and its 2-site cluster extension~\cite{haldar-laad-hassan}.  We 
will only need a single input from CDMFT.  The Hamiltonian is
 
\be
H_{FKM}=-t\sum_{<i,j>}(c_{i}^{\dag}c_{j}+h.c) + U\sum_{i}n_{i,c}n_{i,d} -\mu\sum_{i}n_{i,c}
\ee
and describes a band of dispersive $c$-fermions interacting via a local 
potential $U$ with dispersionless $d$-fermions.  As discussed before, while 
it deceptively resembles a non-interacting disorder model, the annealed as 
opposed to quenched disorder makes it an interacting problem: the competition between $U$ and $t$ in the FKM is slaved to the random $d$ fermion configuration.

%\vspace{0.5cm}
 
\section{Formulation of the Self-Consistent Theory}

%\vspace{0.5cm}

  In order to construct an analytic approach to the MIT from 2-site CDMFT 
  results, we will need to recapitulate some results we will need.  
CDMFT yields cluster-local propagators, $G({\bf K},\omega)$ and self-energies, 
$\Sigma({\bf K},\omega)$, or their equivalents in the ``S,P'' basis of a 2-site
 cluster, $G_{S(P)}(\omega),\Sigma_{S(P)}(\omega)$, upon making a ``cluster-to-orbital'' identification: $S\rightarrow {\bf K}=(0,0)$, and 
$P\rightarrow {\bf K}=(\pi,\pi)$.  For each of the $S,P$ channels, 
the self-consistency condition of cellular-DMFT leads to the equivalence 
of the Ward identity for the lattice model with that for the corresponding two-site
``impurity'' model (most conveniently expressed in the $S,P$-basis as

\be
\Sigma_{a}(\nu+\omega) - \Sigma_{a}(\nu) = T\sum_{\nu'}\gamma_{\nu\nu'\omega}^{a}[g_{a}(\nu'+\omega)-g_{a}(\nu')]
\ee
where $a=S,P$ and $\gamma^{a}$ is the two-particle self-energy of the impurity model.  Using DMFT selfconsistency, we have $g_{a}(\nu)=(1/N)\sum_{k}G_{a}({\bf k},\nu)$, where $G_{a}({\bf k},\nu)$ is the cDMFT propagator obtained earlier~\cite{haldar-laad-hassan}.  Then,
\begin{widetext}
\be
\Sigma_{a}(\nu+\omega)-\Sigma_{a}(\nu) = \frac{T}{N}\sum_{k',\nu'}\gamma_{\nu\nu'\omega}^{a}G_{k',\nu'}^{a}G_{k',\nu'+\omega}^{a}[i\omega -(\Sigma_{a}(\nu'+\omega)-\Sigma_{a}(\nu'))]
\ee
and thus,

\be
\frac{\Sigma_{a}(\nu+\omega)-\Sigma_{a}(\nu)}{i\omega} =\frac{T}{N}\sum_{k',\nu'}\gamma_{\nu\nu'\omega}^{a}G_{k',\nu'}^{a}G_{k',\nu'+\omega}^{a}[1-\frac{\Sigma_{a}(\nu'+\omega)-\Sigma_{a}(\nu')}{i\omega}]
\ee

  On the other hand, for any arbitrary ${\bf q}$, the Bethe-Salpeter eqn for the
  two-particle vertex reads

\be
F_{\nu\nu'}^{a}({\bf q},\omega)=\gamma_{\nu\nu'\omega}^{a} + \frac{T}{N}\sum_{k'',\nu''}\gamma_{\nu\nu''\omega}^{a}G_{k'',\nu''}^{a}G_{k''+q'',\nu''+\omega}^{a}F_{\nu''\nu'}^{a}({\bf q},\omega)
\ee

  We multiply this eqn by $G_{k'}^{a}G_{k'+q}^{a}$ and sum the result over ${\bf k}',\nu'$ to obtain

\be
\frac{T}{N}\sum_{k',\nu'}G_{k',\nu'}^{a}G_{k'+q,\nu'+\omega}^{a}F_{\nu\nu'}^{a}({\bf q},\omega)=\frac{T}{N}\sum_{k',\nu'}\gamma_{\nu\nu'\omega}^{a}G_{k'\nu'}^{a}G_{k',\nu'+\omega}^{a}[1+\frac{T}{N}\sum_{k'',\nu''}G_{k'',\nu''}^{a}G_{k''+q'',\nu''+\omega}^{a}F_{\nu''\nu'}^{a}({\bf q},\omega)]
\ee

Comparing eqns (7) and (9), we see that they actually represent the {\it same} eqn.  Hence,

\be
-\frac{\Sigma_{a}(\nu+\omega)-\Sigma_{a}(\nu)}{i\omega} = \frac{T}{N}\sum_{k'',\nu''}G_{k'',\nu''}^{a}G_{k''+q'',\nu''+\omega}^{a}F_{\nu'',\nu'}^{a}({\bf q},\omega)
\ee

\end{widetext}

  Hence, the three-leg vertex actually varies as the inverse of the quasiparticle residue:

\be
\Lambda({\bf q},\omega) = 1 - \frac{\Sigma_{a}(\nu+\omega)-\Sigma_{a}(\nu)}{i\omega} \rightarrow z_{a}^{-1}(\omega)
\ee

We now use our earlier cDMFT result for the FKM, where we found that Im$\Sigma_{c}(\omega)\simeq |\omega|^{\alpha}$ with $\alpha=1/3$ at the
\begin{figure}
\includegraphics[width=1.\columnwidth , height= 
1.\columnwidth]{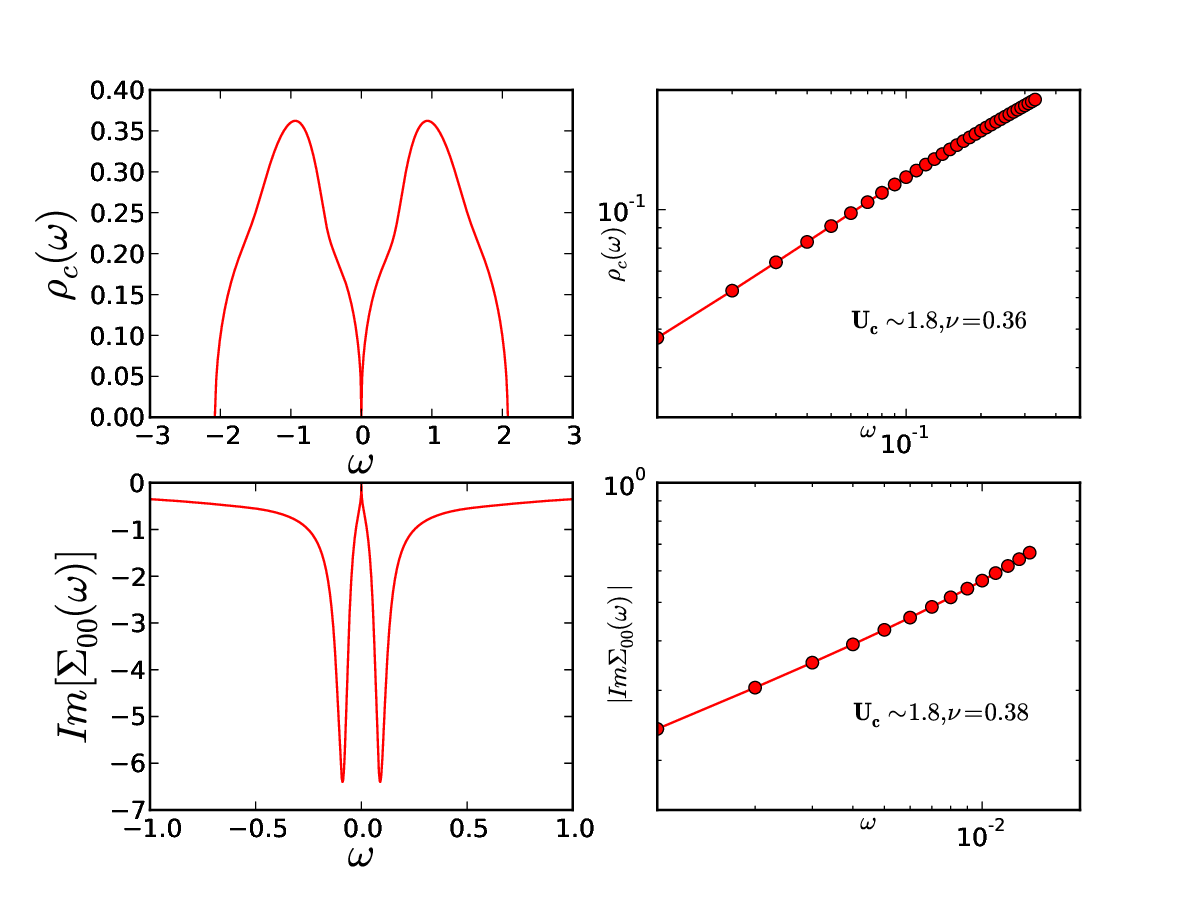} 
\caption{(Color online) Exponent of $\rho_{00}(\omega)$ and $Im\Sigma_{00}(\omega)$ closed to the Fermi energy at critical value of U with symmetric alloy}
\label{fig:fig11}
\end{figure}
Mott-like MIT as shown in Fig.~\ref{fig:fig11}. This gives $z_{c}(\omega)\simeq \omega^{1-\alpha}$, leading to an infra-red divergence of the three-leg vertex, 
$\Lambda_{c}(\omega)\simeq \omega^{-(1-\alpha)}$.  Thus, the singular fermion self-energies at the MIT within cDMFT directly lead to 
infra-red singular vertices, and to drastic modification of the charge fluctuation spectrum.  Explicitly, 

\be
\chi_{ch}(q,\omega) \simeq \frac{1}{\omega.z_{c}^{-1}(\omega)+iD_{0}q^{2}}\simeq [\omega^{\alpha}+iD_{0}q^{2}]^{-1}
\ee  
Thus, the pole structure of the diffusion propagator in the diffusive metal (or in a correlated Landau Fermi liquid with a finite Landau quasiparticle residue,
$0 < z_{a} <1$, is supplanted by a branch-point singularity at the 
``Mott'' quantum critical point in our case.  In a Landau Fermi liquid (LFL), 
$z_{a}(\omega)=z$, a constant in the infra-red, leading to regular diffusion modes.  We can also interpret the above eqn in terms of an anomalous diffusivity at the MIT (see below).     

  It is remarkable that this result is reminiscent of those found in the pioneering 
  two-loop renormalization-group (RG) analysis of the disordered and interacting electron liquid in $D=2$~\cite{punnoose-finkelstein}.  However, the spatially 
  long-range correlations crucial in the RG work are irrelevant in our 
  ``high-$D$'' approach.  Instead, the non-perturbative cluster correlations 
  leading to singular Fermi liquidity underpin our findings above.  
  
  Following Vollhardt and W\"olfle~\cite{VW}, our results can also be recast in terms of a continuum hydrodynamical description of the interacting fluid.  The 
  dynamical density fluctuation propagator, expressed in terms of the dynamic 
  diffusivity, $D(q,\omega)$, or equivalently in terms of the generalized current relaxation kernel, $M(q,\omega)$ (the ``memory function'') is

\be
\chi(q,\omega)=\frac{-e^{2}n/m}{\omega^{2}+\omega M(q,\omega)-(e^{2}n/m\chi(q,0))}
\ee
with 

\be
M(q,\omega)=i\frac{n}{m\chi(q,0)}D^{-1}(q,\omega)
\ee
These equations apply very generally to {\it any} continuum quantum fluid.  From the above, the memory function, $M(q,\omega)$ exhibits anomalous behavior at the ``Mott'' QCP,

\be
M(q,\omega)=M(0,\omega)\simeq D^{-1}(0,\omega)\simeq \omega^{-(1-2\alpha)}
\ee
of the infra-red singular branch-cut form (see below, in the discussion on optical conductivity).

%\vspace{1.0cm}

\section{Effects of Magnetic Impurities}

%\vspace{0.5cm}

  Near the QCP, the $c$-fermion local densty-of-states (LDOS) acquires a 
non-analytic form in two-site CDMFT~\cite{haldar-laad-hassan}, 
$\rho_{c}(\omega)\simeq C|\omega|^{\alpha}$ with $\alpha=1/3$ as shown in Fig.~\ref{fig:fig11}.  If we now 
dope a small number of ``magnetic impurities'', the system at the QCP is 
nothing other than the lattice version of a power-law pseudogap Kondo impurity model,

\be
H_{K}=\sum_{k}(\epsilon_{k}+\Sigma(K,\omega))c_{k\sigma}^{\dag}c_{k\sigma} +
J\sum_{i,\sigma,\sigma'}{\bf S}_{i}.c_{i\sigma}^{\dag}{\bf \sigma}_{\sigma\sigma'}c_{i\sigma'}
\ee
where the ``conduction'' electrons themselves can undergo a band-splitting type
of continuous MIT.

  A lot is known about such a model~\cite{Ingersent-Si}.
In this case, the local moments can be quenched by $c$-fermions only for
$J>J_{c}$.  For $H_{K}$, the local moments are asymptotically decoupled from
conduction electrons and, with a disordered distribution of local moments,
the system is a random, quantum critical spin liquid as long as $J<J_{c}$.  In this situation, intersite correlations between the localized moments will either result in
$(i)$ RKKY-interaction induced random singlet phase,
 quenching the extensive degeneracy of the partially quenched quantum-critical state.  For $J>J_{c}$ in the metallic phase, we expect the 
local moments to be quenched by the itinerant fermion spin density, driving
heavy Landau-Fermi liquid formation.  Precisely at the associated QCP, since we
have particle-hole symmetry, a symmetric critical point (SCP) separates 
the weak and strong coupling regimes for $0<\alpha<0.375$~\cite{ChenJayaprakash}.  
Near the QCP, the singular component of the impurity free energy 
is~\cite{Ingersent-Si}

\be
F_{imp}=T.f(|J-J_{c}|/T^{a}, |h|/T^{b})
\ee
where $h$ is the local magnetic field on the impurity.  The critical behavior 
of the local moment amplitude and the spin susceptibility for $\alpha=1/3$  
follows as $M_{loc}(J<J_{c},T=0=h) \simeq (J_{c}-J)^{\beta}$, $M_{loc}(J=J_{c},T=0)\simeq |h|^{1/\delta}$ and $\chi_{loc}(J=J_{c})=C_{1}T^{-x}$, with 
$\beta=(1-b)/a=0.3548, \delta=b/(1-b)\simeq 14, x=(2b-1)=0.86$.  Correspondingly, the imaginary part of the local dynamical spin suscetibility exhibits
$\omega/T$ scaling:

\be
Im\chi_{loc}(\omega,T)=\frac{\tau_{0}^{\nu}sin(\pi\nu/2)}{2(2\pi T)^{1-\nu}}B(\frac{\nu}{2}-i\frac{\omega}{2\pi T},1-\nu)
\ee
whence Im$\chi_{loc}(J=J_{c},\omega,T=0)=C'|\omega|^{-\nu}$sgn$\omega$, with 
$\nu=x$.  Since the exponents describing the ower-law fall-offs in 
quantum-critical charge and spin dynamics are distinct, this critical phase 
hosts a high-$D$ {\it spin-charge separation}!  This can be tested in Kondo alloys near 
the metal-insulator transition (see below), and is indicative of emergence of a high-$D$
random spin liquid in disordered local moment alloys.  It is rather interesting that this 
form 

\be
T^{1-\nu}Im\chi_{loc}(\omega,T)=\frac{\tau_{0}^{\nu}sin(\pi\nu/2)}{2(2\pi)^{1-\nu}}B(\frac{\nu}{2}-i\frac{\omega}{2\pi T},1-\nu)
\ee
with $B(x,y)=\frac{\Gamma(x)\Gamma(y)}{\Gamma(x+y)}$ and $\Gamma(z)$ the gamma function, is similar to the ``strange metal'' form~\cite{SYK}.  In our case, this holds down to $T=0$ only 
at the Mott-like QCP.  This could constitute a novel, multifermion pair glue for onset of unconventional superconductivity in Kondo alloys around such a QCP.  

%\vspace{1.0cm}

\section{Implications for Experiments}

%\vspace{0.5cm}

  These novel findings have remarkable consequences for physical responses at
  the ``Mott'' QCP.  Since there is no scale other than temperature, $T$, in the quantum-critical regime of a quasi-local QCP, we infer, upon replacing $\omega$ by $T$, that $z_{a}(T)\simeq T^{-(1-\alpha)}$ (see, for e.g,~\cite{punnoose-finkelstein}.  This has several interesting consequences for experiment:

$(i)$ the specific heat at constant volume is given by $C_{v}(T)=z_{c}(T)T\simeq T^{\alpha}$, implying a critically diverging $\gamma$-co-efficient,
$\gamma_{el}\simeq T^{-(1-\alpha)}$.

$(ii)$ The spin susceptibility of the liquid near the QCP is $\chi_{s}(T) \simeq T^{-x}$, while 
 the spin diffusion constant, $D_{s}$, scales anomalously to zero.

$(iii)$ thus, remarkably, the Wilson ratio, $R_{W}=\chi_{s}/\gamma_{el}$, is enhanced  and diverges like 
$T^{1-\alpha-x}$ when $(\alpha +x) >1$ in the quantum-critical regime.  If $(\alpha +x)<1$, $R_{W}(T)$ is suppressed relative to its 
non-interacting value and vanishes at $T=0$.  Vanishing of $R_{W}(T\rightarrow 0)$ has
been actually proposed from numerical studies for various spin liquid models~\cite{prelovsek-spin-liquids}.

$(iv)$ the electron spin resonance (ESR) line-width scales with the spin susceptibility, {\it i.e}, the ratio~\cite{Sachdev}

\be
\frac{\Delta H_{1/2}(T)}{\Delta H_{1/2}(0)}=\frac{D_{s}(0)}{D_{s}(T)}\simeq \chi_{s}(T)\simeq T^{-x}
\ee
increases in an anomalous way as $T\rightarrow 0$ near the QCP.

\begin{figure}
\includegraphics[width=1.\columnwidth, height= 
1.\columnwidth]{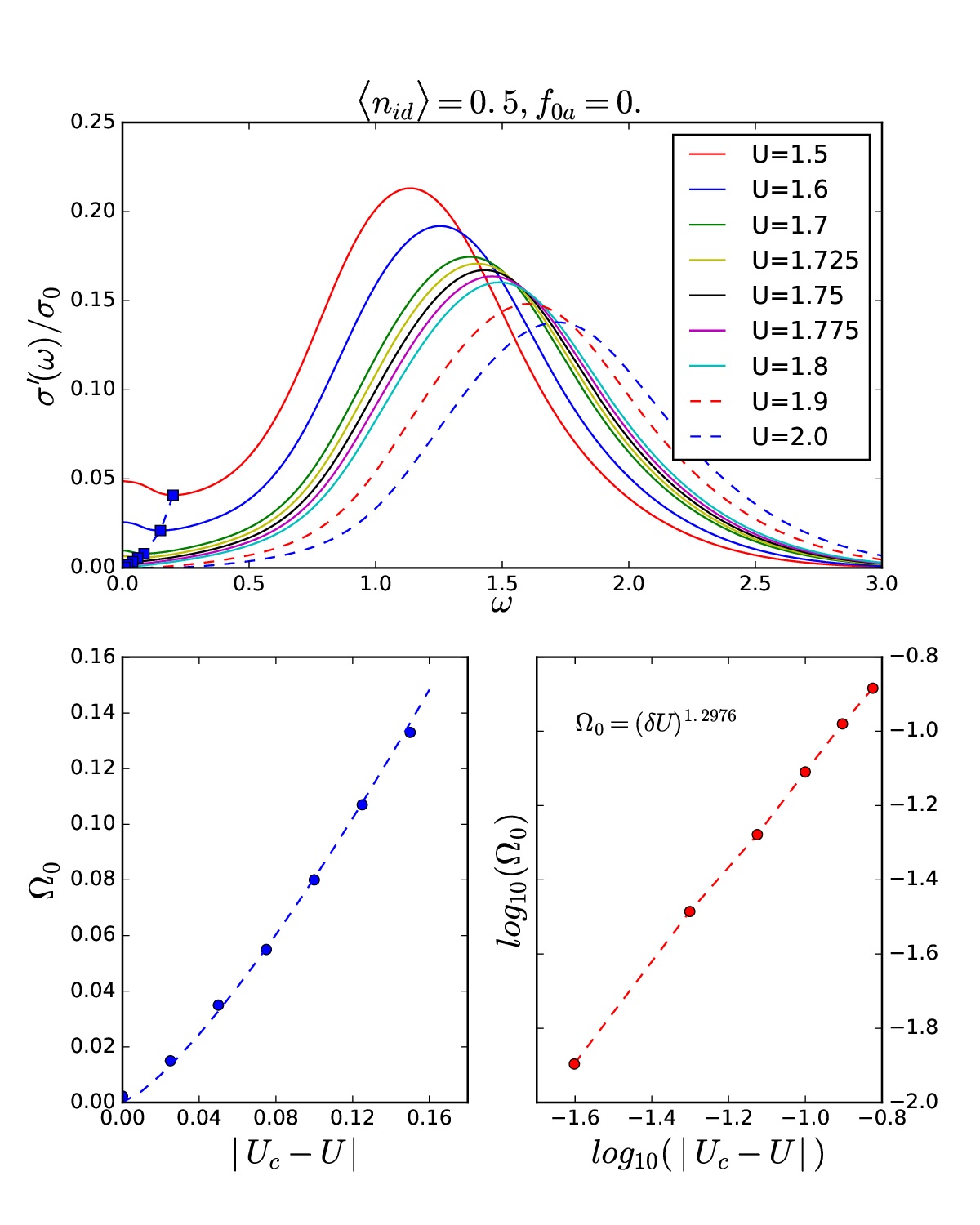} 
\caption{(Color online) Optical Conductivity of the completely random ($f_{0\alpha}=0$) FKM within two-site CDMFT, showing its evolution with $U$ at temperature $T\rightarrow 0$.  The MIT occurs at $U_{c}=1.8$.  Blue symbols show how an emergent 
scale, $\Omega_{0}(U)$, associated with a smooth crossover between metallic and insulating states, collapses at the Mott transition ($U=1.8$) as $(\delta U)^{\nu}$ with $\nu=1.29$, close to $4/3$
(see text)}
\label{fig:fig1}
\end{figure}

\begin{figure}
\includegraphics[width=1.\columnwidth , height= 
1.\columnwidth]{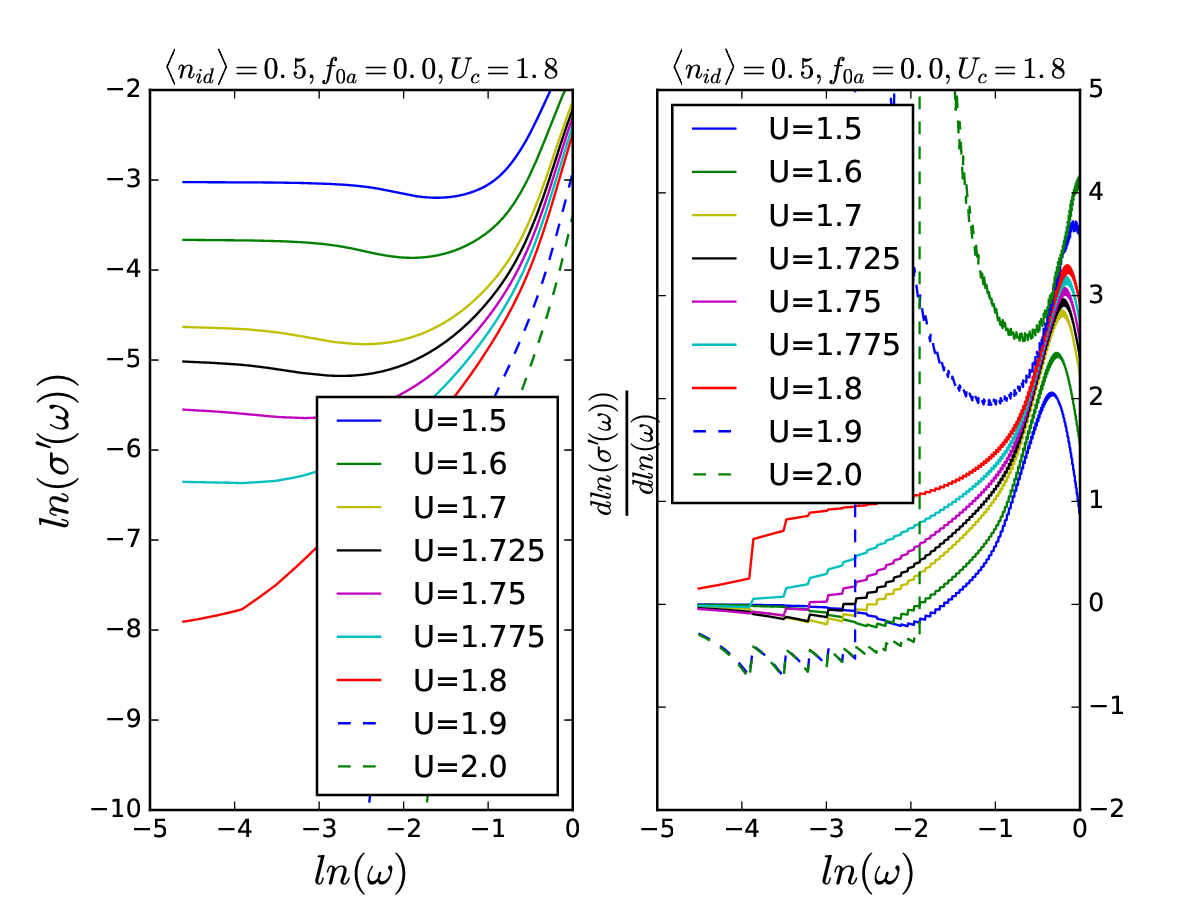} 
\caption{(Color online) Real part of the optical conductivity versus frequency, $\omega$, plotted on a log-log scale.  The crossover from the $dc$ limit at very low frequencies in the bad metal 
to UDR around ln$(\omega)\simeq -3$ close to the Mott QCP (for $U\simeq 1.8$) is clearly seen.  The $dc$ limit contribution progressively vanishes in the proximity of the Mott QCP, and UDR emerges in the quantum critical region associated with the continuous MIT.}
\label{fig:fig5}
\end{figure}

$(v)$ the optical conductivity of the fluid is related to $\chi(q,\omega)$ by

\be
\sigma_{xx}(0,\omega)=-e^{2}.Lim_{q\rightarrow 0}(\frac{\omega}{q^{2}})\chi''(q,\omega)
\ee
This holds because, due to the continuity equation, the correlation functions of charge and current are connected to each other.
Substituting $\chi(q,\omega)$, we find that $\sigma_{xx}(0,\omega)\simeq e^{2}\rho(0)D_{0}\omega^{1-2\alpha}$ with $\alpha =1/3$.  The dynamic diffusivity, $D(0,\omega)$ has the same behavior in 
this regime.  In Fig.~\ref{fig:fig1}, Fig.~\ref{fig:fig5}, we show our earlier results for the optical response of the FKM across the MIT~\cite{PRB_Thoss_2019,PRB_Zonda_2020}.


  Away from the Mott QCP, this behavior is cut off by the finite gap in
the insulator, and by a finite $\omega$-independent but incoherent metallic 
contribution in the metal.  Using our result $\sigma_{xx}(\omega=0,\delta U)\simeq (\delta U)^{\nu}$ with $\nu=4/3$, we can construct an interpolative form

\be
\chi_{ch}(q,\omega,\delta U)\simeq \frac{1}{\omega z^{-1}(\omega)+iD_{ch}(q^{2}+\xi^{-2})}
\ee
with $\xi^{-1}\simeq (U-U_{c})^{\nu}$ and a formally non-vanishing $z$ away from the QCP in the metal (where the $c$-fermion DOS exhibits a pseudogap).  We 
then recover a finite $\sigma_{xx}(0,\omega)$ and diffusive charge fluctuations
throughout the metallic phase, with subdiffusion emerging {\it only at the QCP}.
 Remarkably, this is a mathematical statement of the 
famed Jonscher's law~\cite{Jonscher}, which has been used widely to interpret a
 wide range of data in disordered systems for a long time.  This form, with 
the same exponent $\alpha=1/3$ is also seen in the VW selfconsistent theory 
of localization, but at the {\it Anderson} transition.  Here, we find that 
this holds at the band-splitting type of metal-insulator transition (on the 
Bethe lattice), where the {\it average} DOS, rather than the {\it typical} DOS, exhibits non-analytic behavior at the MIT. 

\vspace{0.5cm}

$(vi)$ Quite generally,the quasi-elastic scattering dynamical structure factor 
(DSF) on an interacting electron fluid is

\be
S_{ch}(q,\omega)=\frac{1}{\pi}Re[\frac{1}{\omega+D(0,\omega)f(q)}]
\ee

For small $q$, we have $f(q)\simeq q^{2}$, and for 
$|\omega| >> q^{2}D(0,\omega)$,
we find that

\be
S_{ch}(q \rightarrow 0,\omega) \simeq D'(0,\omega)/\omega^{2} \simeq \omega^{-(2-\alpha)}
\ee
This form is well known to describe, for {\it e.g}, Hydrogen diffusion in amorphous metals like Ni$_{24}$Zr$_{76}$~\cite{EPL}.  For our case, such an anomalous DSF near the MIT can be probed in, e.g, inelastic X-ray scattering studies.

$(vii)$ complementary information of the subdiffusive phase obtains from the 
return
probability, $P(r=0,t)$,a quantity that is directly related to the long-time
behavior of the local density fluctuation propagator.  Generally, the Fourier 
transform of the non-local density correlation function is given as

\be
P(q,t>0) =\int_{0}^{\infty} d\omega e^{-i\omega t}\frac{D(q,\omega)q^{2}}{\omega^{2}+(D(q,\omega)q^{2})^{2}}
\ee
At long times ($\omega\rightarrow 0$), we find that $P(r=0,t) \simeq C.t^{-\eta}$ with $0<\eta<1$.  Thus, both, optical conductivity (which measures long-wavelength, low-energy response) and the return probability (which quantifies long-time, local response) exhibit anomalous subdiffusive behavior with exponents 
depending continuously on ``disorder''.  This suggests possible quantum 
Griffiths-like behavior at 
the QCP between a ($T=0$) diffusive metal and a Mott-like insulator.  And we 
find that the subdiffusive metal shrinks to a single point, coinciding with 
the Mott-like QCP.

$(viii)$  Anomalous responses as above must also have remarkable manifestations in 
the short- and long-time relaxation of the electric polarization in response 
to sudden changes (quenches) in an external electric field, which we now 
consider.  This is especially clear upon using the above result for $\sigma(0,\omega)$ in the formulation of Rypdal {\it et al.}~\cite{Rypdal}, which we do now.

  In response 
to a suddenly switched-off external electric field, the subsequent decay of 
polarization is given by 

\be
{\bf P}({\bf r},t)=\int_{-\infty}^{+\infty}\chi_{el}(t-t'){\bf E}({\bf r},t')dt'\ee
  If the field is suddenly switched off at $t=0$, i.e, if
${\bf E}({\bf r},t)={\bf E}({\bf r})\theta(-t)e^{\nu t}$ with $\nu=0^{+}$ the decay of the polarization at times $t>0$ is determined by a relaxation function, $\phi(t)$, defined as ${\bf P}({\bf r},t)=\phi(t){\bf E}({\bf r})$.
We have

\be
\phi(t)=\theta(-t)e^{\nu t}\phi(0) + \theta(t)[\phi(0)-\int_{0}^{t}\chi_{el}(t')dt']
\ee
with $\chi_{el}(t)=-(d\phi/dt)$ and $\phi(0)=\int_{0}^{\infty}\chi_{el}(t')dt'$.  To make progress, we use a
macroscopic formulation, using $\sigma(0,\omega)$ as the sole input from microscopics.  Generally, we have
$\nabla.{\bf E}=4\pi\rho({\bf r},t)$ and $\nabla.{\bf P}=-\rho({\bf r},t)$, whence $\nabla.{\bf D}=0$ with
${\bf D}={\bf E}+4\pi{\bf P}$.  The polarization current is ${\bf J}_{P}=(\partial {\bf P}/\partial t)$.  Then
${\bf J}({\bf r},t)=\int_{0}^{\infty}\sigma(t-t'){\bf E}({\bf r},t')dt'$ with $\sigma(t-t')=\frac{\partial}{\partial t}\chi_{el}(t-t')$.  In Fourier space, $\sigma(\omega)=-i\omega\chi_{el}(\omega)$.  In the critical region,
we thus have $\chi_{el}(\omega)\simeq (\omega)^{-2\alpha}$, which is just a restatement of the well known polarization catastrophe at a continuous MIT.

   Laplace transforming the above equations and noting that $\nabla.{\bf J}=(\partial/\partial t)\nabla.{\bf P}=-{\partial\rho}/{\partial t}$, we arrive at the relations $\nabla.{\bf E}(r,\omega)=4\pi\rho(r,\omega)$,
${\bf J}(r,\omega)=\sigma(\omega){\bf E}(r,\omega)$, and thus $\nabla.{\bf J}(r,\omega)=\sigma(\omega)\nabla.{\bf E}(r,\omega)=4\pi\sigma(\omega)\rho(r,\omega)$, while the continuity equation translates into
$\omega\rho(r,\omega)-\rho(r,0)+\nabla.{\bf J}(r,\omega)=0$.  Inserting $\sigma(\omega)\simeq \omega^{1-2\alpha}$
with $0<\alpha <1$ from $(5)$ above, we find that

\be
\omega\rho(r,\omega) + 4\pi\sigma_{0}\omega^{1-\alpha'}\rho(r,\omega)=\rho(r,0)
\ee
with $\alpha'=2\alpha$.
In the spirit of spatial locality of the response, we separate variables, $\rho(r,\omega)=\phi(\omega)\psi(r)$ to get

\be
\phi(\omega)=\frac{\phi(0)}{\omega+4\pi\sigma_{0}\omega^{1-\alpha'}}
\ee
The time-dependent response is then the inverse Laplace transform

\be
\phi(t)=\frac{1}{2\pi i}\int_{-i\infty}^{+i\infty}\frac{e^{t\omega}}{\omega+\tau^{-\alpha'}\omega^{1-\alpha'}}d\omega
\ee
This is just the Mittag-Leffler function, $E_{\alpha}[-(t/\tau)^{\alpha}]$ and admits a series expansion

\be
E_{\alpha'}[-(t/\tau)^{\alpha'}]=\sum_{p=0}^{\infty}(-1)^{p}\frac{(t/\tau)^{p\alpha'}}{\Gamma(p\alpha'+1)}
\ee
This has very revealing short- and long-time behavior: at short times $t<<\tau$, $E_{\alpha'}[-(t/\tau)^{\alpha'}]\simeq exp[-(t/\tau)^{\alpha'}]$, a stretched exponential dependence, while for long times, $t>>\tau$, $E_{\alpha'}[-(t/\tau)^{\alpha'}]\simeq t^{-\alpha'}$.

   Quite remarkably, experiments on two-dimensional electron gas systems by Popovic {\it et al.}~\cite{Popovic} seem to
show this precise $t$-dependent behavior, both at small and large times.  
This suggests onset of glass-like dynamics in the proximity of the MIT.
Very close to the MIT, we expect this behavior to persist down to very large
times, but to exhibit an eventual crossover to the diffusive form at even longer
 times.  Finite-size numerics may miss this latter crossover due to intrinsic
size constraints.  

     Thus, the above results suggest that glassy behavior (at $T=0$) seems to 
onset precisely at the quantum critical point associated with the MIT, and is linked to onset of subdiffusive dynamics at this Mott-like QCP.
 It is of interest to observe that similar
behavior is encountered in studies of many-body localization (MBL) in $D=1$ spin-$1/2$ 
XXZ models in a random magnetic field~\cite{demler}.  Moreover, similar optical
responses vis-a-vis our results~\cite{Laad-optics-FKM} have been seen in the optical 
response near the MIT in numerics.  Our findings within 
cDMFT unearth an anomalous critical state at $T=0$ {\it only at the QCP}.  
 It turns out that while we are unable to find a 
non-ergodic metal {\it phase} within the present approach for the FKM, 
the Mott-like insulator is a genuine non-ergodic phase: we turn to this aspect 
below.

%\vspace{1.0cm}

\section{LDOS Correlations and Non-Ergodicity at the Metal-Insulator Transition}

%\vspace{0.5cm}

Let $\phi_{\mu}(r_{i})$ and $\epsilon_{\mu}$ denote the eigenstates and eigenvalues of an electron in a disordered, interacting liquid.  We define the two-particle spectral functions~\cite{Lee-Ramakrishnan},
\begin{widetext}
\be
\langle\langle\rho_{E}(r_{i})\rho_{E+\omega}(r_{j})\rangle\rangle^{M} = \frac{1}{\rho(E)}\langle\sum_{\mu\mu'}\phi_{\mu}^{*}(r_{i})\phi_{\mu'}(r_{i})\phi_{\mu'}^{*}(r_{j})\phi_{\mu}(r_{j})\delta(E-\epsilon_{\mu})\delta(E+\omega-\epsilon_{\mu'})\rangle
\ee
and

\be
\langle\langle\rho_{E}(r_{i})\rho_{E+\omega}(r_{j})\rangle\rangle^{L} = \frac{1}{\rho(E)}\langle\sum_{\mu\mu'}|\phi_{\mu}(r_{i})|^{2}|\phi_{\mu'}(r_{j})|^{2}\delta(E-\epsilon_{\mu})\delta(E+\omega-\epsilon_{\mu'})\rangle
\ee

\end{widetext}
with $\rho(E)=\langle\sum_{\mu}|\phi_{\mu}(r_{i})|^{2}\delta(E-\epsilon_{\mu})\rangle$ being the average one-electron DOS.  The above eqs. are correlators of 
the local densities-of-states. Their Fourier components obey the ``sum rules''
$\langle\langle\rho_{E}\rho_{E+\omega}\rangle\rangle_{q=0}=\delta(\omega)$, 
and $\int d\omega \langle\langle\rho(E)\rho(E+\omega)\rangle\rangle_{q}=1$, 
with $\langle\langle\rho(E)\rho(E+\omega)\rangle\rangle_{q} \geq 0$.  These two
correlators coincide at $r_{i}=r_{j}$.

  The important point is that the Fourier transform of the first of the above 
correlators is related to the two-particle Green function of the fermionic 
model via

\be
\langle\langle\rho(E)\rho(E+\omega)\rangle\rangle_{q}^{M} = \frac{1}{\pi\rho(E)}Im[\Phi_{E}^{RA}(q,\omega) - \Phi_{E}^{RR}(q,\omega)]
\ee
where $R(A)$ denote retarded and advanced components, $\Phi_{E}^{RA(R)}(q,\omega) = -(2\pi i)^{-1}\sum_{p,p'}\langle G^{R}(p_{+},p_{+}',E+\omega)G^{A(R)}(p_{-}',p_{-},E)\rangle$, and $p_{\pm}=p\pm (1/2)q$.
  Since only $\Phi^{RA}$ is singular at the MIT, we neglect $\Phi^{RR}$ in the
following.  Using

\be
\Phi_{E}^{RA}(q,\omega) = -\frac{\rho(E)}{\omega + iD(\omega)q^{2}},
\ee  
we can explicitly evaluate the LDOS correlators.  We still need $D(\omega)$ in the diffusive metal, subdiffusive critical region and the Mott-like insulator to
make progress.  

Since $D(\omega)\simeq \sigma_{xx}(0,\omega)$ as shown above, we use the cDMFT
results for the optiical conductivity of the FKM in these three regimes.  We
have

(i) $D(\omega)=D_{0}$ for $|\omega|<< \omega_{c}\simeq \xi^{-1}(U)\simeq (U_{c}-U)^{\nu}$, with $\nu=1.3\simeq 4/3$,

(ii) $D(\omega)=D_{0}\omega^{1-2\alpha}e^{-i\pi(1-2\alpha)/2}$ in the quantum-critical region of the MIT, where $\omega_{c}=0$, and

(iii) $D(\omega)=i\omega\xi^{2}$, just on the insulating side.  Using these equations, the LDOS correlator is evaluated as ($K_{\rho\rho}(q,\omega)=\langle\langle\rho(E)\rho(E+\omega)\rangle\rangle_{q}^{M}$),

\be
K_{\rho\rho}(q,\omega)=(1/\pi)\frac{D_{0}q^{2}}{\omega^{2}+(D_{0}q^{2})^{2}}
\ee
in the diffusive metal, with $D_{0}=D(\omega\rightarrow 0)>0$,

\be
K_{\rho\rho}(q,\omega)=(1/\pi)\frac{D(\omega)q^{2}}{\omega^{2}+(D(\omega)q^{2})^{2}}
\ee
in the quantum-critical state at the MIT, with $D(\omega)\simeq D_{0}\omega^{1-2\alpha}$ and
\begin{widetext}
\be
K_{\rho\rho}(q,\omega)=(1/\pi)(1+q^{2}\xi^{2}(U))^{-1}\delta(\omega)=A(q)\delta(\omega) 
\ee
\end{widetext}
just on the insulating side of the MIT.  

The Fourier transform of $A(q)$, $A(r)=\int d^{3}q A(q)e^{i{\bf q}.{\bf r}}$, is
the inverse participation ratio (IPR), defined to be

\be
A(r)=\frac{1}{\rho(E)}\langle\sum_{\mu}\delta(E-\epsilon_{\mu})|\phi_{\mu}(r)|^{2}|\phi_{\mu}(0)|^{2}\rangle
\ee
That $K_{\rho\rho}(r,0) \simeq \delta(\omega)$ signals non-ergodicity in the 
Mott-like insulator~\cite{PRB_Zonda_2019}.  Remarkably, breakdown of ergodicity in the insulator was
already anticipated and discussed by Anderson in his seminal 1958 paper (Anderson, 1958).  However, this is {\it not} so for the diffusive metal.  At the MIT, however, we find, for $r=0$, 
$d=3$ and small $\omega$, that

\be
\langle\langle\rho_{E}(0)\rho_{E+\omega}(0)\rangle\rangle^{M} \simeq C.\omega^{-\alpha}
\ee
This is a very interesting result, and a qualitatively similar form has 
been found in earlier work dealing with multifractal nature 
of eigenstates at an Anderson transition~\cite{Feigelman}.  We find that 
the Mott-like QCP is also associated with IR-singular LDOS correlations, and 
this may be related to multifractality of eigenstates in our non-Anderson  
MIT~\cite{Radzihovsky,Radzihovsky1}.  As far as we are aware, this anomalous power-law behavior of the local LDOS correlator at a non-Anderson MIT has not been 
noticed before. 

%\vspace{1.0cm}

\section{Quantum Geometry of the Mott Transition:}   On a deeper level, we 
now exploit above results to also study modification 
of polarization fluctuations and the quantum metric tensor across the MIT.  
We begin by considering the electric polarization as a linear response 
to a uniform electric field, $H_{ext}=-EX=-E\sum_{i,\sigma}x_{i}n_{ic\sigma}$.
  The electric polarization, $P$ is then $P=L^{-D}\langle X\rangle=-L^{-D}(\partial H/\partial E)=L^{-D}\partial_{E}H$, where $H=H_{FKM}+H_{ext}$.  
The electric susceptibility is $\chi_{el}=(dP/dE)|_{E=0}=-L^{-D}\partial_{E}^{2}H|_{E=0}$.  This is also expressible as usual by the relation~\cite{baeriswyl}

\be
\chi_{el}=2L^{-D}\sum_{m\neq 0}\frac{|\langle\psi_{0}|X|\psi_{m}\rangle|^{2}}{\epsilon_{m}-\epsilon_{0}}
\ee
with $\epsilon_{m},\epsilon_{0}$ being the energies of eigenstates $\psi_{m},\psi_{0}$.  Obviously, $\langle X\rangle|_{E=0}=0$.  Baeriswyl {\it et al.}~\cite{baeriswyl} define a charge gap $\Delta=(\epsilon_{m}-\epsilon_{0})$ with the lowest value of $\epsilon_{m}$ for which $\langle X^{2}\rangle\neq 0$, giving $\chi_{el}\leq (2/\Delta)L^{-D}\langle\psi_{0}|X^{2}|\psi_{0}\rangle$, with $\langle\psi_{0}|X^{2}|\psi_{0}\rangle$ being the estimate of polarization fluctuations.  Since we have

\be
L^{-D}\langle\psi_{0}|X^{2}|\psi_{0}\rangle=\xi,
\ee
the localization length, it is clear that the static polarization fluctuations exhibit a critical divergence at the MIT, of the form $\langle X^{2}(U)\rangle \simeq (U-U_{c})^{-\nu}$, with $\nu$ the correlation length exponent, equal to
$\nu=4/3$ for the FKM within two-site CDMFT~\cite{haldar-laad-hassan}.

  We are now in a position to address the critical behavior of the quantum 
geometric tensor at the continuous MIT in the FKM.  For a spatially isotropic system,
 the full metric tensor reduces to $Q_{\mu\nu}=g_{\mu\nu}$ with 
$\mu=\nu$~\cite{resta}.  Moreover, since

\be
g_{\mu\nu}=\sum_{m\neq 0}\frac{\langle\psi_{0}(\lambda)|\partial_{\mu}H|\psi_{m}(\lambda)\rangle\langle\psi_{m}(\lambda)|\partial_{\nu}H|\psi_{0}(\lambda)\rangle}{[\epsilon_{m}(\lambda)-\epsilon_{0}(\lambda)]^{2}}
\ee
(where $\lambda$ is a tuning parameter, the metric tensor will exhibit singular behavior when $\Delta =0$.
Specializing to the MIT in our case, we find that

\be
g_{EE}=\sum_{m\neq 0}\frac{\langle\psi_{0}|\partial_{E}H|\psi_{m}\rangle\langle\psi_{m}|\partial_{E}H|\psi_{0}\rangle}{\epsilon_{m}-\epsilon_{0}}
\ee

This function is bounded as $g_{EE}\leq\sum_{m\neq 0}\Delta_{m,0}^{-2}\mid\langle\psi_{0}\mid\partial_{E}H\mid\psi_{m}\rangle\mid^{2}=\Delta^{-2}\sum_{m\neq 0}\mid\langle\psi_{0}\mid\partial_{E}H\mid\psi_{m}\rangle\mid^{2}$.  But this is just equal to
$\Delta^{-2}[\langle\partial_{E}H\partial_{E}H^{\dag}\rangle-|\langle\partial_{E}H\rangle|^{2}]$ .  Since $\langle X\rangle=-\partial_{E}H$,
we find that

\be
g_{EE}=\Delta^{-2}[\langle X^{2}\rangle-\langle X\rangle^{2}]=\frac{\langle X^{2}\rangle}{\Delta^{2}}
\ee
 
     Thus, at the Mott QCP, the QGT diverges because the polarization fluctuations diverge there.  This is a
remarkable result, because the QGT is manifestly a non-Landau like quantity, having no interpretation in 
terms of a local order parameter.  Nevertheless, it satisfies all the criteria for a novel order parameter:
it's {\it inverse} is finite in the insulator, exhibits critical behavior at the QCP, and vanishes in the metal.  It thus turns out to be a non-Landau order 
parameter for the continuous (non-Anderson) MIT in the FKM.  We point out 
that a similar result is known for the Anderson disorder model~\cite{resta}, but our finding establishes it's role as a ``geometric'' order parameter for the simplest example of a continuous Mott transition.  Power spectral density (PSD) of voltage noise across the QCP may be a promising way to quantify this unconventional order parameter (see below)~\cite{Neupert}

%\vspace{1.0cm}

\section{Discussion}

%\vspace{1.0cm}

  We have developed an analytic approach for the continuous MIT in the 
FK model, using a single input from our earlier cDMFT work.  As emphasized at 
the beginning, the FK model is an appropriate effective model when we consider 
the physical case of a random ensemble of charged ``impurities'' in a metallic 
host.  These play a dual role: they scatter the itinerant 
$c$-fermions via a local, Hubbard-like interaction.  Since the $d$-fermions are
 randomly distributed, this is a random FK model.  Now, $(1)$ the $d$-fermions 
present a random, local interaction for  the $c$-fermions, 
$(2)$ they mediate an {\it effective} coulomb interaction between itinerant 
$c$-fermions, and $(3)$ generate a $d$-fermion configuration-dependent hopping 
between $c$-fermions.  Thus, in contrast to the non-interacting Anderson 
disorder model, we are dealing with a model of spinless fermions with 
random hopping {\it and} local interactions, along with (emergent under renormalization) non-local coulomb interactions when inter-site correlations 
within an embedded cluster are considered explicitly.  In a single-site DMFT, 
these features are absent.  

  Such a model falls into the class of those that have acquired importance in 
  the context of many-body localization (MBL) in recent times~\cite{Altshuler,Huse,Huse1,Prelov}.  Models with disorder and interactions have a long history and, in what has
  subsequently been referred to as the 
  MBL context, were described by Anderson and Fleischman~\cite{Anderson-Fleishman,Zonda_2009,Zonda_2012,PRB_Zonda_2019,PRB_Zonda_2020, PRB_Thoss_2019}.  They have been tackled mostly with numerical approaches in $d=1$~\cite{demler}, as well as by high-$d$ cavity approximations~\cite{Logan} in Fock space.  Moreover, since the $d$-fermion mediated $c-c$ 
  interaction in our case is expected to be small compared to the local, random Hubbard-like term, the effective model is precisely a spinless model 
of electrons with random on-site potential, random hopping, and weak non-local 
  interactions, which is precisely the limit where one might have expected MBL-like responses to occur.  Finally, our two-site cluster DMFT is in fact a 
version of the cavity approximation that is exact to $O(1/D)$.  The above findings suggest 
that an anomalous sub-diffusive ``metal'' intervenes {\it only} at the QCP separating the
weakly diffusive metal and the non-ergodic Mott-like insulator.  We find 
that this subdiffusive ``phase'' at finite but low $\omega,T$ progressively shrinks to a single point.  This happens at the Mott-like QCP associated with the transition from a bad metal 
to a Mott-like insulator (at $U_{c}=1.8$ within two-site CDMFT for the FKM).  It hosts 
non-ergodic features {\it above} a low energy scale that shrinks to zero at 
the MIT, at least within this quasi-local theory.

  But we are unable to capture
 multifractal correlations in our approach.  The reason seems to be tied with the 
observation that the $q$-dependence of the static susceptibility, $\chi(q,0)\simeq (q^{2}+\xi^{-2}(\delta U))^{-1}$ at small $q$ 
continues to have a $q^{2}$ form.  Multifractality of eigenstates at a non-Anderson MIT has been investigated by Syzranov {\it et al}~\cite{Radzihovsky,Radzihovsky1}, but has not been investgated using embedded cluster approaches, to our best knowledge.  Nonetheless, we {\it are} able to show that the 
correlator of the LDOS exhibits anomalous branch-point structure at the Mott 
QCP, see above.

\vspace{1.0cm}

  Notwithstanding the inability to capture the ``bad metal'' phase argued first
 by Altshuler~\cite{Altshuler}, our approach does succeed in describing the
 subdiffusive dynamical responses (at low but finite energy) in the very bad metallic phase near the non-ergodic insulator.  These features are 
qualitatively similar to those encountered in 
classical glassy systems and, while the existence of an electron glass phase
preceding the localization transition~\cite{Dobrosavljevic} has 
been proposed within DMFT, dynamical responses in such a phase have not, to our
best knowledge, been considered theoretically.  Moreover, the fate of such
glassy behavior in a situation with {\it annealed} disorder remains unclear.
  We now detail how our results are also consistent with extant numerical 
results~\cite{Tomasi} for dimensions $D>2$.

  To shed light on this aspect, we now analyze the spectral responses of both, the ``light'' $c$-fermions, and the ``infinitely heavy'' $d$-fermions across the MIT (we notice that the FKM is isomorphic to the ``light-heavy'' mixtures that 
  have been considered in the context of the quantum disentangled liquid 
(QDL) state~\cite{MPAFisher}, and our analysis should also apply 
to such systems in $D\geq 3$).  While the FKM maps onto the usual Anderson 
disorder model as far as the $c$-fermions are concerned, the extra ingredient is the (non-trivial) $d$-fermion dynamics in the FKM, arising from annealed as 
opposed to quenched (positional) randomness in the $d$-fermion configuration.

     Let us focus on the $c$-fermion sector to begin with.  We begin with the
 density correlator, $G(r,t)=(1/V)$Re$\sum_{i}Tr[\rho_{0}n_{i+r}(t)n_{i}(0)]$,
 where $\rho_{0}$ is the ``initial state'' density matrix, taken equal to unity
 for temperature $T=\infty$.  The time-dependent mean-square displacement~\cite{Luitz},

\be
\langle x^{2}(t)\rangle =\sum_{r}r^{2}G(r,t)
\ee
varies like $t$ in the diffusive, and like $t^{2/z}$ with $z=z(U/t)>2$ in the 
subdiffusive regime.  For a thermal initial state, this is related to the 
charge diffusivity,$D_{ch}(\omega)$, by
\begin{widetext}
\be
D_{ch}(\omega) = -\omega^{2}\int_{0}^{\infty}dt e^{i\omega t}\langle x^{2}(t)\rangle = \omega^{2}\int_{0}^{\infty}dt e^{i\omega t}|t|^{2/z}\simeq \omega^{1-(2/z)}=\omega^{1-2\alpha}
\ee
\end{widetext}
with $\alpha=1/z<1$, fully consistent with our finding before.  This is also 
consistent with numerics in $D=1$.  In two-site CDMFT, we studied the 
optical conductivity, $\sigma_{xx}(0,\omega)$ across the MIT in the FKM.  

\vspace{1.0cm}

  A direct comparison of our results with those of Prelovsek et al.~\cite{prelovsek-spin-liquids,Prelovsek} reveals 
  close similarities in the optical response across the MIT.  We refer the reader to our previous CDMFT results (Haldar-Laad-Hassan (2019)~\cite{Laad-optics-FKM}).  It is clear that: (1) $\sigma_{xx}(0,\omega)$ is constant at 
  low energies in the incoherent metal, below the low energy scale, $\Omega\simeq (\delta U)^{z\nu}$ with $z=1, \nu=4/3$ that vanishes at the QCP.  This scale 
corresponds to the crossover from bad-metallic to an insulating behavior, and 
continuously reduces as the MIT is approached, vanishing just at the MIT as 
$\Omega(\delta U)\simeq (U_{c}-U)^{4/3}$.  Exactly at the ``Mott'' MIT, we 
find that $\sigma_{xx}(0,\omega) \simeq \omega^{\alpha}$ with $\alpha=1/3$: 
this implies $\alpha=1-(2/z)<1$ with $z=3$, and onset of non-ergodicity 
precisely at the MIT.  Throughout the metallic phase, 
$\sigma_{xx}(0,\omega)=C(\delta U)$ for $\omega<\Omega(\Delta U)$.  This implies that the metallic phase is ergodic all the way upto the QCP, where non-ergodicity sets in.

\vspace{0.5cm}

Our results can also be used to investigate noise power spectra (NPS) across the MIT. At the MIT, we expect dramatic increase in noise due to divergent fluctuations. On either side of the MIT, one would still expect enhanced noise, but that this growth should be cut off by the finite correlation length, $\xi \simeq |U-U_{c}|^{-\nu}$. This would be expected to lead to a “flattening out” in the NPS at low frequencies (long times). The Wiener-Khintchine theorem tells us that the power spectral density (PSD) of a time series $X(t)$ is essentially the Fourier transform of the time-correlator
of the fluctuations, $\delta X(t)=(X(t) - X(0))$. At very low T, these must be the quantum fluctuations of the electronic density in our case. The above arguments tell us that the PSD at the QCP must fall off like $S_{\rho\rho}(\omega)\simeq 1/\omega^{K}$ with $K>1$.
Close to $U_{c}$, we write~\cite{Yu}

\be
S_{\rho\rho}(\omega)=\Omega^{-\gamma}F(\omega/\Omega^{z\nu})
\ee

where $\Omega = \mid U-U_{c}\mid^{z\nu}$, $F$ a scaling function, $\nu$ the correlation length exponent, $\xi\simeq \mid U-U_{c}\mid^{-\nu}$, and $z$ the dynamical critical exponent. $\gamma$ is the critical exponent associated with the divergent charge susceptibility.  

     But at the QCP, $\chi_{\rho\rho}(\omega)$ must be independent of 
$\Omega$, so that we must have $K= \gamma/z\nu$. Hence, Re$\chi_{\rho\rho}(\omega)\simeq \omega^{-\gamma/z\nu}$ and, using the
Kramers-Krönig relation, Im$\chi_{\rho\rho}(\omega)\simeq \omega^{-\gamma/z\nu}$ for $\omega >0$ as well. The fluctuation-dissipation theorem then implies the
PSD given by

\be
S_{\rho\rho}(\omega)=\frac{4k_{B}T}{\omega}Im\chi_{\rho\rho}(\omega)\simeq \omega^{-(1+\gamma/z\nu)}
\ee

Now, we use the fact that $\gamma=1$ for the critical exponent of the ``order parameter'' susceptibility (that is associated with instability to Wigner CDW order) in single-site~\cite{freericks} as well as two-site cDMFT. This leads to $S_{\rho\rho}(\omega)\simeq \omega^{-7/4}$, with an
exponent $K>1$ as promised before. The physical origin of this power-law enhancement is the anomalously vanishing $z(\omega)\simeq \omega^{\alpha}$ and the large scale transfer of dynamical spectral weight from high to low energy 
upon approach to the Mott-like QCP.  In the metallic
phase, since $\sigma_{xx}(\omega)\simeq const$ below a low frequency
scale $\Omega_{c}(U_{c}-U)$ which vanishes at the QCP, we find that 
$\epsilon(0,\omega)\simeq \omega^{-1}$ for $\omega < \Omega_{c}$, yielding a 
``$1/f$'' type behavior of the voltage noise PSD.  The crossover from $f^{-K}$
($K>1$) to $1/f$ behavior should thus be governed by the scale 
$\Omega_{c}(U)\simeq (U_{c}-U)^{4/3}$.  It is very interesting to note that 
a PSD of voltage noise in the vicinity of a {\it correlation-driven} first-order
 MI transition in NdNiO$_{3}$ on suitable substrates exhibits an anomalous
exponent $K=1.2-1.3$~\cite{Arup}.  While we obtain $K=1.75$, 
it is possible that use of larger clusters will reduce $K$.  Though we get sizable enhancement of noise at the QCP, more is needed to make any
comparison with real systems: first, the MIT in the FKM is 
continuous, while that in NdNiO$_{3}$ is clearly first-order with 
hysteresis.  Second, the insulating phase has antiferromagnetic order, while our
 result holds for the case without any such conventional order across the MIT.  On the other hand, the metallic phase above $T_{MI}\simeq 120-130$~K is 
clearly a bad metal without Landau quasiparticles, likely violating the Mott-Ioffe-Regel criterion ($k_{F}l <1$).  A nanoscopic, intrinsic electronic 
phase separation (EPS) close to $T_{MI}$ is invoked to rationalize enhanced PSD 
at low energy~\cite{Arup}.  Adopting an effective FKM-like view for the proposed EPS scenario, 
one may visualize the insulating patches with an annealed, random distribution 
of $d$-electrons (antiferromagnetically correlated inside insulating droplets 
but uncorrelated between different droplets) between insulating patches, and 
the metallic islands by a set of $c$-fermions in the ``metallic'' regions in a suitably coarse-grained FKM picture.  

     Anomalous voltage noise PSD in the vicinity of the Mott transition has further, interesting consequences in the
context of breakdown of ergodicity in mixtures of mobile fermions and infinitely heavy entities.  A percolative conduction mechanism of the MIT, as suggested 
by our finding of $\nu=4/3$ and enhanced noise PSD implies an inhomogeneous 
current distribution.  Careful work on VO$_{2}$ and NdNiO$_{3}$ shows that 
in the $T$-window of the MIT ($T_{1}<T<T_{2}$), the Nyquist noise, 
which measures the electronic temperature $T_{e}$, is sizably {\it enhanced} above the equilibrium lattice temperature ($T=T_{l}$).  If we consider 
the lattice as a bath to which electrons are coupled, having 
$T_{e}\neq T_{l}$ means that the electron subsystem has lost thermal equilibrium with the (phonon bath) lattice.  It is interesting to inquire whether this is
related to the light- heavy- particle ``decoupling'' and breakdown 
of ergodicity in light-heavy fermion mixtures in the ``quantum disentangled
liquid''~\cite{MPAFisher} context.  In the experimental cases above, it may suggest such a ``decoupling'' between localized and itinerant fermions in the insulating and metallic patches in a proposed EPS picture.   
  
  Thus, onset of subdiffusive dynamics in the critical region of the continuous 
  MIT manifests in anomalous deviation from the ubiquitous $1/f$-law in the
  quantum-critical region, and is a measurable effect that can 
quantify subdiffusion.
  
\vspace{1.0cm}

     We can also specify quantum criticality at the MIT as follows.  The FKM
     reduces to the Anderson disorder problem as far as the conduction electron
     response is concerned.  However, what is {\it very} interesting is that we
     find such features at the ``Mott-like'' band-splitting MIT, as opposed to 
     the Anderson MIT~\cite{Ohtsuki} in numerics.  Explicitly, our previous
     cDMFT work gives the $dc$ conductivity, $\sigma(\delta U)=e^{2}/\hbar\xi=(\delta U)^{\nu}$ with $\nu=1.3$, the correlation length exponent, in excellent accord with numerics for the orthogonal 
class~\cite{Ohtsuki}.  As regards $\langle x^{2}(t)\rangle$, we follow Ohtsuki
{\it et al.} and assume the scaling form,

\be
\langle x^{2}(t)\rangle = Ct^{k_{1}}F[(U_{c}-U)t^{k_{2}}]
\ee
This scaling form is motivated by the observation of $\langle x^{2}(t)\rangle \simeq t^{\alpha}$, with $\alpha <1$ at the MIT above.  On the one hand, in the
diffusive metal, one must have $\langle x^{2}(t)\rangle =2Dt\simeq \sigma(\delta U).t \simeq (U_{c}-U)^{s}t$, (here, $D$ is the diffusion constant, not to be confused with the spatial dimension, $D=3$) so that we must have $F(x)\simeq x^{s}$, with 
$s=(D-2)\nu=\nu$ for $D=3$.  Thus, we get $\langle x^{2}(t)\rangle =C.t^{k_{1}+sk_{2}}(U_{c}-U)^{s}$, implying the relation $k_{1}+sk_{2}=1$.  On the other hand, in the insulator, diffusion stops as soon as $\langle x^{2}(t)\rangle \leq \xi^{2} \simeq (U-U_{c})^{-2\nu}$, so that we have $F(x)\simeq (-x)^{-2\nu}$ for large $(-x)$, and implies $k_{1}-2\nu k_{2}=0$.  Hence, we find $k_{1}=2/3$ and
$k_{2}=1/(3\nu)=1/4$, and thus $\langle x^{2}(t)\rangle \simeq t^{2/3}$, which is also in excellent accord with numerics~\cite{Ohtsuki}.  Using the relation, 
$D_{ch}(0,\omega)=-\omega^{2}\int dt e^{i\omega t}\langle x^{2}(t)\rangle$, we
get $D_{ch}(0,\omega)\simeq \sigma_{xx}(0,\omega)\simeq \omega^{1/3}$ in $d=3$.
This seems to be in excellent accord with the two-site CDMFT results for the FKM, where $\sigma_{xx}(0,\omega)\simeq \omega^{\alpha}$ with $\alpha=0.34\simeq 1/3$ for $U=1.775$, coinciding with $U_{c}=1.8$ within numerical error.

\vspace{0.5cm}

     It is intriguing that these findings closely resemble those obtained from
     direct numerical studies of the Anderson transition in disordered electronic systems in $D=3$ dimensions.  We can only offer the following plausibility argument which,
by no means, should be considered as a ``derivation''.  
     To study Anderson localization (AL) in quantum cluster theories, it is 
     necessary to study the geometrically averaged Green function in a 
     typical-medium DMFT
     (TM-DMFT), rather than the algebraically averaged GF we have studied.  
     TM-DMFT can
     capture AL, even in single-site DMFT.  This has indeed been done by many
     groups by now, so we appeal to the most recent work of Terletska {\it et al.}~\cite{Terletska-Raja} for $D=3$ (replacing the actual $D=3$ 
     band DOS by a semi-elliptic form is a
     good approximation in $D=3$, since both feature a square-root behavior near the band edge).  A direct perusal of their results shows up interesting 
     features germane 
     to our query above.  Specifically, it is found that the {\it typical} DOS,
     $\rho_{typ}(\omega)$ decreases faster than the average DOS as a function 
     of $U/W$, establishing an Anderson insulating phase before the band-splitting ``Hubbard-like'' MIT.  Importantly, however, $\rho_{typ}(\omega)$ at low energy near the Anderson MIT seems to vary (with frequency) in a manner similar to that exhibited by the average DOS.  In the single site limit ($N_{c}=1$), both 
     transitions 
     seem to occur at almost the same $U_{c}$: for small cluster sizes beyond DMFT, this conclusion still seems to hold, and a clean distinction only obtains
     for cluster sizes larger than $N_{c}=8$.  Thus, if we use $\rho_{typ}$ to
     investigate the quantum criticality at the Anderson MIT within our 
two-site cluster, we would expect, at best, a very small difference 
in the critical $(U/W)$s separating the Anderson and band-splitting Mott-like 
transitions.  If it turns out that $\rho_{typ}(\omega)$ varies as 
$\omega^{1/3}$ at the Anderson MIT (this seems to be the case, but remains to 
be verified explicitly in extant results of Terletska {\it et al.}),
 the features we have studied here would all obtain at the Anderson MIT, simply  by replacing the average DOS by the typical DOS.  Observation of the {\it same} exponents in numerical work of Ohtsuki {\it et al.} suggests such an 
 eventuality, and it would be interesting to verify it.

\vspace{0.5cm}

   Furthermore, in the context of glassy behavior indicated by subdiffusive 
   responses, 
   it is also of (also experimental) interest to analyze the dynamical
 structure factor (DSF) of density fluctuations.  Quite generally, the DSF is

\be
F(q,t)=Re(Tr[\rho_{0}n_{q}(t)n_{-q}(0)])
\ee
with $n_{q}(t)=\frac{1}{\sqrt{V}}\sum_{i}n_{i}(t)e^{iq.R_{i}}$.  This is the
quantum analogue of the coherent inelastic scattering function in classical 
glasses.  Quite generally, the DSF for disordered matter is

\be
F(q,t)=Re(Tr[\rho_{0}n_{q}(t)n_{-q}(0)]) \simeq exp(-Dq^{2}\langle x^{2}(t)\rangle)
\ee
with $n_{q}(t)=\frac{1}{\sqrt{V}}\sum_{i}n_{i}(t)e^{iq.R_{i}}$.  This is the
quantum analogue of the coherent inelastic scattering function in classical 
glasses.  In the diffusive regime, we thus expect that $F(q,t)\simeq exp(-D_{0}q^{2}t)$ with $D_{0}=$Lim$_{\omega\rightarrow 0}D(\omega)$.  However, at the QCP, a 
qualitative change occurs in the above: we find that the DSF exhibits a stretched exponential form,
 $F(q,t)\simeq$ exp$(-D_{0}q^{2}t^{1-\alpha})$, a Kohlrausch-Williams-Watts (KWW) form characteristic of what is found in spin glasses and supercooled liquids.  
 This explicitly links subdiffusion with electronic glassiness, and both to a 
percolative metal-insulator transition (recall that the correlation length exponent we find is $\nu\simeq 4/3$ in cDMFT, and $\nu=1.2$ in single-site DMFT).  

     Finding of KWW relaxation in a two-site CDMFT framework for the FKM is remarkable.  We believe that this occurs because, close to the MIT, the dynamics is
{\it hierarchically constrained}~\cite{Palmer1984,Palmer1984_err}.  We present the following
heuristic argument: If the ``return to equilibrium'' involves $c$-fermion
 hopping in a background of many sequential, correlated activation steps, 
it must be strongly constrained by the occupancy of (cluster) sites onto 
which it is to hop: if both cluster sites have no $d$-fermion, a $c$-fermion 
can readily hop onto either of them.  If both are occupied by a $d$-fermion, 
$c$-fermion hopping is blocked due to the appreciable ($U\simeq O(W)$) energy 
cost needed to occupy either cluster site.  If one of the two cluster sites is 
occupied by a $d$-fermion (this configuration is two-fold degenerate on a 
two-site cluster), a $c$-fermion can of course hop onto the other, empty 
cluster site.  Clearly, the $c$-fermion hopping is slaved to the localized $d$-fermion 
configuration, and this (local) constraint persists over wide timescales 
$t>>\hbar/W, \hbar/J_{z}$.  The $c$-fermion dynamics is in a sense dynamically 
frustrated by the $d$-fermion ``disorder''.  

     Only when the $d$-fermion configuration changes appreciably can the 
character of $c$-fermion dynamics change.  Over longer and longer 
time scales, all (time) scales associated with such annealing 
of the $d$-fermion ``disorder'' distribution must, in the long-time 
($t\rightarrow\infty$) limit, constrain and determine the fate of the long-time
$c$-fermion dynamics.  This is especially so very close to the MIT, where
the interference between these long time scales determines the dynamics.  The 
state of the slow $d$-fermion degrees of freedom will, as exemplified in the 
two-site cluster example above, non-trivially influence the ``faster'' 
$c$-fermion dynamics.  Upon disorder averaging, the system effectively 
``passes through'' a large number of annealed disorder configurations, and hence, through a wide 
range of relaxation times enroute to equilibrium.  
But as the MIT is approached, this gets progressively harder, because the 
increasing blocking of $c$-fermion hopping lengthens the ``faster'' timescale, 
to infinity at the MIT, where the diffusion stops.  Ultimately, this is what leads to onset 
of non-ergodicity and KWW relaxation at the QCP. 
   
%\vspace{0.5cm}

   It is very noteworthy that our finding of subdiffusive non-ergodic behavior 
   {\it only} at the quantum critical point between an ergodic metal and a 
non-ergodic insulator is fully consistent with very recent numerical studies~\cite{Tomasi}, with the important difference that we find these features at 
the QCP associated with a {\it band-splitting} type of a ``Mott-like'' MIT.  
This accord may imply that an exact-to-$O(1/D)$ treatment of the random 
FKM may suffice for all dimensions $D>2$.  Clearly, however, such small cluster sizes as we have used here ($N_{c}=2$) is not adequate for the Anderson-MI 
transition~\cite{Terletska-Raja}.  On the other hand, the case of {\it annealed}, as opposed to quenched disorder remains to be investigated for larger 
cluster sizes within TM-DMFT.  We mention, however, that large scale Monte 
Carlo simulations for the $D=2$ FKM finds three {\it insulating} 
phases~\cite{Antipov}: a CDW at low $T$, an {\it Anderson} insulator for small 
$U/t$ above $T_{cdw}$, and a Mott-like insulator for larger $U/t$ above 
$T_{cdw}$, consistent with folklore.  In our random FKM, we do not expect CDW order.  TM-CDMFT results indicate
 that the threshold for Anderson localization may be driven to lower $U/t$ upon
 consideration of larger cluster sizes, but it is challenging to do this for 
cluster sizes large enough to obtain an Anderson insulator in $D=2$ for any 
$U/t$.  Thus, we expect our results in this paper to hold for $D=3$,
where even single-site DMFT seems to capture many features of the Mott-like 
MIT~\cite{freericks} in $D=3$.

\vspace{1.0cm} 

  The above analysis can also be applied to the spin sector in the physical
  situation where a finite concentration, $x_{s}$, of magnetic impurities are 
  doped in the disordered system.  As we have seen, the disordered magnetic
  system at the MIT can be modeled as a ``pseudogap Kondo-RKKY'' system: it is 
thus a ``disordered Kondo alloy''.  At the quantum critical point, 
$U=U_{c}=1.8$, the ``conduction electron'' ($c$-fermion) DOS varies like 
$\rho_{c}(\omega)=C|\omega|^{1/3}$.  A disordered quantum critical spin-liquid-
like state characterized by a local quantum critical form of dynamic spin 
correlations with $\omega/T$ scaling and fractional exponents results when 
the local Kondo coupling, $J_{K}< J_{K}^{(c)}$.  As discussed before, the 
dynamical spin susceptibility then reads

\be
\chi''(q,\omega)\simeq \frac{1}{\omega.z_{s}^{-1}(\omega)+iD_{0,s}q^{2}}
\ee
where $z_{s}\simeq \omega^{1-\nu_{s}}$ and $D_{0,s}$ is the bare spin-diffusivity.  
This has precisely the same form as the dynamical charge susceptibility but,
as found before, with a spin relaxation that is distinct from relaxation of charge: a specific realization of ``high-$D$'' spin-charge separation in 
Kondo alloys.  Hence, we expect subdiffusive spin relaxation and glassiness in 
the quantum critical region associated with the metal-insulator transition.  In
 particular, we expect anomalously slow NMR and ESR relaxation rates, and that 
the long-time dynamical spin susceptibility,

\be
F_{s}(q,t)=Re(Tr[\rho_{0}S^{z}_{q}(t)S^{z}_{-q}(0)])
\ee

must again show a KWW-like relaxation, $F_{s}(q,t)\simeq$ exp$(-D_{0,s}q^{2}t^{1-\nu_{s}})$, in the quantum critical region.  It is quite interesting to recollect 
that anomalously slow NMR relaxation was in fact one of the main motivations 
for the first proposal of Anderson localization itself~\cite{Anderson}.  It is 
rather interesting that the band-splitting Mott-like MIT also shows such 
features.

%\vspace{1.0cm}

\section{"Heavy" $d$-fermion Dynamics Near the MI Transition}.

%\vspace{0.5cm}

     In the FKM, the $d$-state is distinct from a quenched, random potential.
Reinstating the spin degrees of freedom, it may be viewed as an idealization 
of a situation where a subset of orbital states ($d$), Mott localized by strong electronic correlations co-exist with metallic wide-band ($c$) states.  
This actually occurs in microscopic DMFT studies for multi-orbital
 Hubbard models in suitable parameter ranges.  In this case, the term 
$U\sum_{i}n_{i,d}n_{i,c}$ leads to a range of additional exotica.

  The main difference between the FKM and the usual binary-alloy Anderson 
  disorder model is that the dynamics of the $d$-fermions in the FKM is 
non-trivial.  As far as the $c$-fermions are concerned, the FKM maps onto the 
binary-alloy Anderson disorder model.  We have detailed the spectral 
responses in the $c$-fermion sector in detail up to this point.  However, 
the FKM possesses a {\it local} U$(1)$ gauge symmetry: it is 
known~\cite{Georges} (ignoring spin: adding spin degrees of freedom leads 
to additional, very interesting aspects that are out of scope of our focus here)
 that the ground states with and without a $d$-fermion on any site 
$i$ ($n_{i,d}=1,0$) are degenerate, since $[n_{i,d},H_{FK}]=0$ 
for all $i$.  A ``sudden'' change in the $n_{i,d}$ from $0$ to $1$ acts as a 
``sudden local quench'' for the $c$-fermions, leading to the venerated orthogonality catastrophe and infra-red singular branch-point structure of the local $d$-fermion and local excitonic 
$(c_{i}^{\dag}d_{i}+h.c)$ propagators in the metallic phase within DMFT.  This is reminiscent 
of ``strange metallicity'' in a very different context~\cite{AndersonHTSC}.  Explicitly, one finds,

\be
\rho_{dd}(\omega)=\int dt e^{i\omega t}Im \langle d_{i}(t);d_{i}^{\dag}(0)\rangle\simeq |\omega|^{-(1-\eta)}
\ee
with $\eta=(\delta/\pi)^{2}$ and $\delta=$tan$^{-1}(U/W)$ with $W$ the non-interacting $c$-fermion band-width.  And the local excitonic correlator reads

\be
Im\chi_{cd}(\omega)=\int dt e^{i\omega t}\langle [c_{i}^{\dag}d_{i}(t);d_{i}^{\dag}c_{i}(0)]\rangle \simeq |\omega|^{-(2\eta-\eta^{2})}
\ee
These forms hold qualitatively in cDMFT as well, as long as long as the system 
is metallic, and these infra-red singularities are quenched by the gap opening 
in the insulator.  It is interesting to inquire what happens precisely at 
the MIT, where $\rho_{c}(\omega\simeq E_{F}=0) \simeq |\omega|^{1/3}$ in both 
DMFT and two-site cDMFT.  We expect the above feature to survive in the
metallic phase in our two-site cDMFT.  However, 
precisely at the MIT, one expects a non-trivial change in the $d$-fermion 
response (physically, this corresponds to the spectral response of a suddenly 
created ``core hole'').  Generally, the spectral function of a ``suddenly'' swtiched-on core 
hole in a metal follows from the linked cluster expansion~\cite{Mueller-Hartmann} as

\be
S_{h}(\omega)=\frac{1}{2\pi}\int_{-\infty}^{\infty}dt e^{i\omega t} exp[V_{h}^{2}\int_{0}^{\infty} dE Im\chi_{ph}(E)\frac{e^{-iEt}-1}{E^{2}}]
\ee
with $V_{h}$ the core hole potential (here, $V_{h}=U$).  As long as $Im\chi_{ph}(E)\simeq E$, we have $S_{h}(\omega)\simeq \omega^{-\eta}$ as above. However, 
this undergoes a qualitative change as soon as $Im\chi_{ph}(E)\simeq E^{1-\kappa}$ with $0<\kappa <1$.  One gets~\cite{haldar-laad-hassan}, where we find $\kappa\simeq 0.457$)

\be
S_{h}(\omega)=\frac{V_{h}^{2}}{E_{F}}(\frac{E_{F}}{\omega})^{1+\kappa} exp[-\pi V_{h}^{4}(\frac{E_{F}}{\omega})^{2(1-\kappa)}]
\ee

This corresponds to a very unusual response, 

\be
\rho_{h}(t)|_{t\rightarrow\infty} \simeq e^{-t^{1-\kappa}}
\ee
of the stretched exponential form at long times, at the MIT.  At this point, it is of interest to notice that similar behavior is deduced for the Anderson MIT
by Kettemann~\cite{Kettemann}: He finds that the {\it typical} overlap between a Fermi liquid 
without and with an added (local) potential impurity vanishes at long times, but differently in 
the metal (where it decays like a power-law) and at the QCP (where it decays like a {\it stretched exponential}).  This is precisely what we obtain.  However, our result is for a Mott-like MIT, but
$\kappa$ may bear a relation to multifractal intensity
correlations at a non-Anderson disorder-driven MIT.  This form is again 
reminiscent of the Kohlrausch-Williams-Watts (KWW) relaxation encountered in spin glasses, and is characteristic of onset of non-ergodic behavior.  
It is satisfying that this accords with the onset of non-ergodicity and 
glassiness at the QCP derived earlier from consideration of the return 
probability.
  
\vspace{0.5cm}

     Physically, we propose that knocking out a localized $d$-fermion with a 
high-energy X-ray and probing the long-time relaxation of the resulting 
``core-hole'' may thus constitute an additional diagnostic for onset of 
non-ergodicity at the MIT in the FKM.  It would be very interesting to analyze 
the extant Mn-$2p$ core-level photoemission (PES) spectra in manganites~\cite{DDS} at hole dopings where a paramagnetic insulator-to-ferromagnetic metal 
transition around $T_{c}^{FM}$ occurs, to unearth the intermediate- as well as 
the long-time behavior of the core-hole propagator in both phases.  A more 
complicated, but experimentally ``cleaner'' diagnostic may be one that can 
probe similar changes in local ``excitonic'' response at long times across 
the MIT:  our analysis above predicts a stretched-exponential decay of the 
time-dependent photoluminescence after ``suddenly switching-off'' an external 
optical probe.  It is very interesting that a Fermi edge singularity in the 
photoluminescence lineshape has been observed quite awhile ago by Skolnick {\it et al}~\cite{Skolnick} in InGaAs-InP quantum wells.  Interestingly, it is seen in samples with disorder-induced hole localization, but not in purer samples, 
implicating singularities associated with incipient excitonic effects in a system with co-existing itinerant and localized states.  Whether
the long-time decay of photoluminescence under a sudden quench could exhibit
power-law (in metallic) or stretched-exponential fall-off (at the MIT) is an 
interesting issue that, in our opinion, deserves more study.  

     Finally, it is interesting to notice that the infra-red divergence of
the local excitonic correlator in the FKM tells us that while excitons will 
form (since $U$ is an attractive interaction between a $c$-electron and $d$-hole) for {\it any} $U/t$, the {\it local} $U(1)$ gauge invariance of $H_{FK}$ 
forbids their condensation as a result of Elitzur's theorem~\cite{Barma}.
  Thus, one may characterize the metal as a dynamically fluctuating 
{\it quantum critical} preformed excitonic liquid, and the Mott-like
 insulator as a random ``solid'' of excitons.  In such a situation, any 
relevant perturbation, for example a direct $d-c$ hybridization term, 
$V\sum_{i}(c_{i}^{\dag}d_{i}+H.C)$, will immediately elevates the local 
U$(1)$ invariance to a global U$(1)$ invariance, escaping the constraints 
of Elitzur's theorem and leading to a Bose-Einstein condensate (BEC) of 
pre-formed excitons (see, for example, Batista~\cite{Batista,Batista_err} in this specific 
context).  In this case, the insulator would be a Kondo insulator,
 characterized by a finite $\Delta_{exc}=<c_{i}^{\dag}d_{i}>$, i.e, a mixed valent excitonic BEC.  If the interband hybridization connects states with opposite parities, this insulator will be an electronic ferroelectric.  
Once this happens, the singularities above will be quenched, due to 
gap opening in the excitonic insulator, and due to an infra-red relevant recoil
of the $d$-fermion ($V$ relevant at one-fermion level) in the metal. 

\vspace{1.0cm}

  Our findings also provide a potential well analogy (PWA)~\cite{Economou}  description of the MIT in the FK model.  We observe that $m^{2}(U)=\simeq \xi^{-2}(U)\simeq |U-U_{c}|^{2\nu}$, with $\nu=1.32\simeq 4/3$
from CDMFT~\cite{haldar-laad-hassan}.  This implies bound state formation in 
the insulator (finite $\xi$), it's progressive weakening as the MIT is 
approached, and it's eventual dissociation at the MIT, where $\xi$ diverges.
It is endearing that the ``effective potential'', given by $V_{eff}(r)=v^{2}$Re$\chi(r,0)$, with $\chi(r,0)$ the spatial Fourier transform of $\chi(q,0)\simeq (q^{2}+m^{2}(\delta U))^{-1}$ obtained before in this work,
neatly encapsulates the renormalization of the bare interaction in the metal.  And $v\simeq U^{2}$ in the weak-coupling and $v\simeq J_{z}^{2}$ in the strong coupling limits of $H_{FKM}$
(eq.(1)). 
In fact, the $e^{-m(\delta U)r}/r$ factor is precisely associated with 
the ``screened effective interaction'' with $\xi=m^{-1}(\delta U)$ in the metal.  Notice that this implies that the potential well depth, $m(\delta U)\simeq (U-U_{c})^{\nu}=\xi^{-1}$,
with $\nu=4/3$.  Since we find a critical exponent $\nu=4/3$ characteristic 
of percolation, we suggest that the emergent picture is that of electrons 
trapped in potential wells associated with blocking of percolative pathways in 
the insulator.  In the metal, an above-critical kinetic energy supplies enough 
energy to destabilize the bound state, establishing a percolation pathway for 
incoherent conduction.

\vspace{1.0cm}

  We argue that our picture should be applicable in an effective model framework to a diverse range of real situations.  

\vspace{0.5cm}

     As a first example where our conclusions may be relevant, we notice 
that an effective FK model of itinerant, spinless (due to strong Hund coupling)
$e_{g}$-band fermions, coupled to quasi-static Jahn-Teller distortions (upon
integrating out phonons in the limit of strong electron-phonon coupling) has
been introduced and extensively discussed for CMR manganites~\cite{Ramakrishnan}.  In this picture, one has either a local JT distortion or none, depending upon the fraction of sites where the lone $e_{g}$ electron 
in, say, LaMnO$_{3}$, is not, or is, removed by doping.  The remainder of the $e_{g}$ electrons can hop from JT-distorted sites to undistorted ones, dragging 
their associated local JT distortion along as they hop.  This implies that the 
JT disorder is annealed rather than quenched, and hence, a clear mapping to a
random FK model obtains.  There is strong evidence that the field-induced MIT and CMR in manganites is percolative in nature~\cite{Mydosh}, in qualitative accord with our view.
In manganites, in this view, $e_{g}$ carriers are localized in potential wells 
that are linked to a random but annealed distribution of quasi-static JT 
distortions, and an external magnetic field tilts the balance between localization and itinerance in favor of incoherent conduction via percolative paths.

   It would also be of interest to study charge and spin dynamics near the doping-induced insulator-metal transition(s) in diluted magnetic semiconductors, such as Ga$_{1-x}$Mn$_{x}$As in the context of our finding of anomalously slow, subdiffusive charge and spin dynamics as above.  Finally, the
   Holstein model, where electrons interact with a dispersionless Einstein
   phonon mode, can also be mapped to an effective Falicov-Kimball model with a {\it continuous} distribution of quasi-static disorder when electron-phonon coupling is strong.  It is quite interesting that such models exhibit 
   the famous Mooij correlations~\cite{Ciuchi}.  In fact, precisely similar 
   Mooij correlations (via $(d\rho/dT)<0$ over an extended range in $T$ in the pseudogap 
   region in cDMFT) also appear in our previous cDMFT~\cite{haldar-laad-hassan} work which, given the above-mentioned mapping, is not surprising.  All our conclusions can be 
   directly applied to this case as well.  Such problems have been tackled with ab-initio numerical methods in previous studies~\cite{Majumdar}, and it is encouraging to observe
   that bad-metallic resistivity, crossover from insulating to bad-metallic behavior,
   pseudogap features in optics, etc, are also observed there.  It would be interesting
   to see 
   whether Mott-like quantum criticality and subdiffusion are hidden in those
   approaches, which are tailored to treat larger cluster sizes without, however, the ability to access the thermodynamic limit.  It would be interesting to 
   inquire whether these numerically intensive travelling-cluster approaches can be integrated into cluster-DMFT approaches.  This remains work for the future.

   In fact, our approach should work in any context where itinerant fermions are
  (arbitrarily strongly) coupled to and (static or annealed) Ising variables.  This might
  have a bearing on the ``Quantum disentangled liquid'' (QDL) phase~\cite{MPAFisher} in a model of itinerant fermions coupled to infinitely massive fermions (this is fully isomorphic to the FKM).  
In $D=3$, our approach shows that a non-ergodic, subdiffusive QDL state emerges only at the QCP separating an ergodic metal from a non-ergodic insulator.  Another particularly interesting case is that of a ($Z_{2}$) Ising gauge theory coupled to fermionic ``matter''~\cite{Moessner}, which is rigorously mappable onto 
the spinless FKM.  Applications to the intensively studied superconductor-insulator transition~\cite{Trivedi,NbN} should also be of interest.  We hope 
to investigate some of the above issues in future.

\vspace{1.0cm}

{\bf Acknowledgements}  One of us (M.S. Laad) wishes to thank Professor J. K. Freericks for helpful suggestions, and we thank S. R. Hassan for past collaboration on related topics.

%\section{Figures attached from the optical conductivity paper}

%\newpage
%\newpage
%\newpage
%\begin{thebibliography}{54}

%\end{thebibliography}

% --- Bibliography Section ---

%\bibliography{myreferences} % name of your .bib file without the .bib extension
\end{document}